\documentclass[12pt,twoside]{article}
\usepackage{amsmath,amsfonts}
\usepackage{array}
\usepackage{hyperref}
\usepackage{textcomp}
\usepackage{stfloats}
\usepackage{svg}
\usepackage{url}
\usepackage{booktabs}
\usepackage{verbatim}
\usepackage{balance}
\usepackage{graphicx}
\graphicspath{{./pictures_arxiv/}}
\DeclareGraphicsExtensions{.pdf,.jpeg,.png,.jpg,.eps}
\usepackage[separate-uncertainty=true,binary-units=true]{siunitx}
\usepackage{geometry}
\usepackage{scalerel}
\usepackage{tikz}
\usetikzlibrary{svg.path}
\usepackage{makecell}
\usepackage[caption=false,font=normalsize,labelfont=sf,textfont=sf]{subfig}
\usepackage[section]{placeins}   %
\usepackage[belowskip=20pt,aboveskip=10pt]{caption}
\usepackage{lineno}

\hypersetup{ colorlinks,
	citecolor=blue,
	filecolor=blue,
	linkcolor=blue,
	urlcolor=blue
}

\definecolor{orcidlogocol}{HTML}{A6CE39}
\tikzset{
	orcidlogo/.pic={
		\fill[orcidlogocol] svg{M256,128c0,70.7-57.3,128-128,128C57.3,256,0,198.7,0,128C0,57.3,57.3,0,128,0C198.7,0,256,57.3,256,128z};
		\fill[white] svg{M86.3,186.2H70.9V79.1h15.4v48.4V186.2z}
		svg{M108.9,79.1h41.6c39.6,0,57,28.3,57,53.6c0,27.5-21.5,53.6-56.8,53.6h-41.8V79.1z M124.3,172.4h24.5c34.9,0,42.9-26.5,42.9-39.7c0-21.5-13.7-39.7-43.7-39.7h-23.7V172.4z}
		svg{M88.7,56.8c0,5.5-4.5,10.1-10.1,10.1c-5.6,0-10.1-4.6-10.1-10.1c0-5.6,4.5-10.1,10.1-10.1C84.2,46.7,88.7,51.3,88.7,56.8z};
	}
}
\newcommand{\orcidicon}[1]{\href{https://orcid.org/#1}{\mbox{\scalerel*{
				\begin{tikzpicture}[yscale=-1,transform shape]
				\pic{orcidlogo};
				\end{tikzpicture}
			}{|}}}}

\newcommand{\cen}[1]{\begin{center} #1 \end{center}}

\begin{document}

\cen{\sf {\Large {\bfseries Development and validation of a measurement-driven inter crystal scatter recovery algorithm with in-system calibration } \\  
		\vspace*{10mm}
		Katrin Herweg$^{1}$\orcidicon{0000-0002-2011-7869}, Volkmar Schulz$^{1,2,3,4\dagger}$\orcidicon{0000-0003-1341-9356}, David Schug$^{1,2}$\orcidicon{0000-0002-5154-8303}} \\
	$^1$Department of Physics of Molecular Imaging Systems, Institute for Experimental Molecular Imaging, RWTH Aachen University, Aachen, Germany; $^2$Hyperion Hybrid Imaging Systems GmbH, Aachen, Germany; $^3$Physics Institute III B, RWTH Aachen University, Aachen, Germany
	\vspace{5mm}\\
}
\cen{\pagenumbering{roman}
\setcounter{page}{1}
\pagestyle{plain}
Corresponding Author: Katrin Herweg, Volkmar Schulz. Email: katrin.herweg@pmi.rwth-aachen.de, volkmar.schulz@pmi.rwth-aachen.de\\
}

\begin{abstract}
\noindent\textbf{Background:}
In PET a high percentage of gamma photons being detected undergo Compton scattering in the scintillator. Scintillator blocks are often built from  optically isolated crystals. Depending on the angle of incidence and the scintillator geometry this might lead to inter crystal scatter (ICS) events, where energy is deposited in two or more crystals in the detector, which common positioning and reconstruction algorithms cannot resolve. Therefore, ICS events worsen the spatial resolution and the signal-to-noise ratio in the reconstructed image.\\
\textbf{Purpose:}
We want to address this challenge by recovering individual crystals from ICS events with their corresponding energy deposits. This information could ultimately be fed into an image reconstruction framework.\\
\textbf{Methods:}
In this work, we established an algorithm based on a detector that couples a readout channel to each crystal (one-to-one coupling), which combines a measurement-driven calibration and a fitting routine to achieve the recovery of crystal interactions from measured light patterns. Using Geant4 simulations, we validated and optimized this approach by comparing the recovered events to the simulation ground truth.\\
\textbf{Results:} 
We showed that, with the best performing algorithm versions, all correct crystals could be identified for \SIrange{95}{97}{\percent} of the simulated events and the crystal energies as well as the event energy sum could be recovered adequately. For the event energy sum a deviation of less than \SI{5}{\percent} could be achieved for \SI{96}{\percent} of all events.\\
\textbf{Conclusion:}
Overall, the developed ICS recovery algorithm was successfully validated for one-to-one coupled detector. Future application for other detector configurations should be possible and will be investigated. Additionally, using the new crystal interaction information to determine the most likely first interaction crystal is being examined to improve efficiency and signal-to-noise ratio in the PET reconstruction.

\end{abstract}

\section{Introduction}
\label{sec:introduction}
In positron emission tomography (PET) a high percentage of gamma photons being detected undergo Compton scattering in the scintillator \cite{Zhang2019}. Depending on the angle of incidence and the scintillator geometry this might lead to inter crystal scatter (ICS) events, where energy is deposited in two or more crystals in the detector. ICS is a long-known challenge for PET \cite{Comanor1996,Levin1997,Rafecas2001,Rafecas2003a}. It results in wrongly assigned lines of responses (LORs) due to falsely positioned events, thereby worsening the spatial resolution \cite{Shao1996,Gillam2014,Surti2018}. In regular data processing only one crystal ID is passed to the reconstruction. Since it is not possible to determine the first-interaction ID of an ICS event with complete certainty, it seems advantageous to consider also first-interaction probabilities and other crystal IDs in the reconstruction \cite{Gillam2012a}. One option to deal with ICS, when it can be resolved, is to reject these events before the reconstruction to avoid a degradation of the spatial resolution \cite{Gross-Weege2016, Ritzer2017}. However, this reduces the detector sensitivity noticeably. At this point in time, a positive effect on the image quality in clinical scans has not been shown yet. To tackle this challenge several methods have been proposed including neural network approaches \cite{Wu2020,Lee2021} and signal multiplexing \cite{Shim2023}. We use a measurement-driven algorithm inspired by Gross-Weege et al (2016) \cite{Gross-Weege2016} and similar to that described in Lee et al (2018) \cite{Lee2018}, using a linear combination of known light patterns to estimate the measured light pattern. Within the algorithm we aim to very cleanly determine the single crystal interaction (SCI) response of the detector, which then includes all optical and readout effects. This makes the achieved SCI light spread matrix (SCI-LSM) unique for each calibrated detector. What differentiates our approach from that of Lee et al (2018) \cite{Lee2018} is the direct application to measurement data, the usage of different positioning patterns in the calibration process 
and its integration in a processing framework capable of online and post processing \cite{Goldschmidt2013,Goldschmidt2016}. In the end we aim to make crystal interactions accessible for an existing reconstruction framework, in order to utilize all the information available. For now, we focus on a clinical one-to-one coupled detector geometry and validate the algorithm with Geant4 simulations. This simulation models the interaction of an incoming gamma photon with the scintillator without describing the scintillation process itself or the detection of the optical photons by the photosensor. Energy deposits in crystals can be transformed into measured light patterns using the SCI-LSM. The results of the validation are presented here.

\section{Materials}
\label{sec:materials}
The proposed algorithm is based on a measurement-driven calibration. In this work, we use a one-to-one coupled PET detector to get the necessary calibration data. The detector consisted of a $12 \times 12$ LSO array (Tianle Photonics Co. LTD.), which was subdivided into $2 \times 2$ crystal sub elements by a double layer of ESR reflector (see Fig. \ref{fig:crystal_array}). Each of the crystal pixels in a sub element had a size of $\SI{3.92}{\milli\meter} \times \SI{3.92}{\milli\meter} \times \SI{16}{\milli\meter}$. No glue was used inside of the array to optimize the light transport. As photo-sensors, $36$ digital silicon photomultipliers (dSiPMs, DPC3200-22 by PDPC) in a $6 \times 6$ arrangement were employed, resulting in $12 \times 12$ readout channels with a pitch of $\SI{4}{\milli\meter}$ matching the crystal pitch. One crystal sub group is located on a dSiPM that outputs photon values for all four channels when the readout condition is met \cite{Tabacchini2014}. The readout was accomplished by the Hyperion III platform \cite{Weissler2020}. It should be noted that the application of the method to other sensor technologies, such as analogue SiPMs read out with ASICs \cite{Nadig2021,Nadig2019}, is viable, as the algorithm takes no prior knowledge of the detector into account except for the measured light pattern. Measurements were conducted using a $^{22}$Na source, which irradiated the whole crystal array at a detector temperature of \SI{17}{\degreeCelsius}. We employed DPC trigger scheme 2 with validation network 8-OR (see Tabacchini et al (2014) \cite{Tabacchini2014} for a more detailed description) at an overvoltage of \SI{3}{\volt}.\\
\subsection{Geant4 simulation}
\begin{figure}[h]
	\centering
 	\subfloat[]{\includegraphics[width=0.4\textwidth]{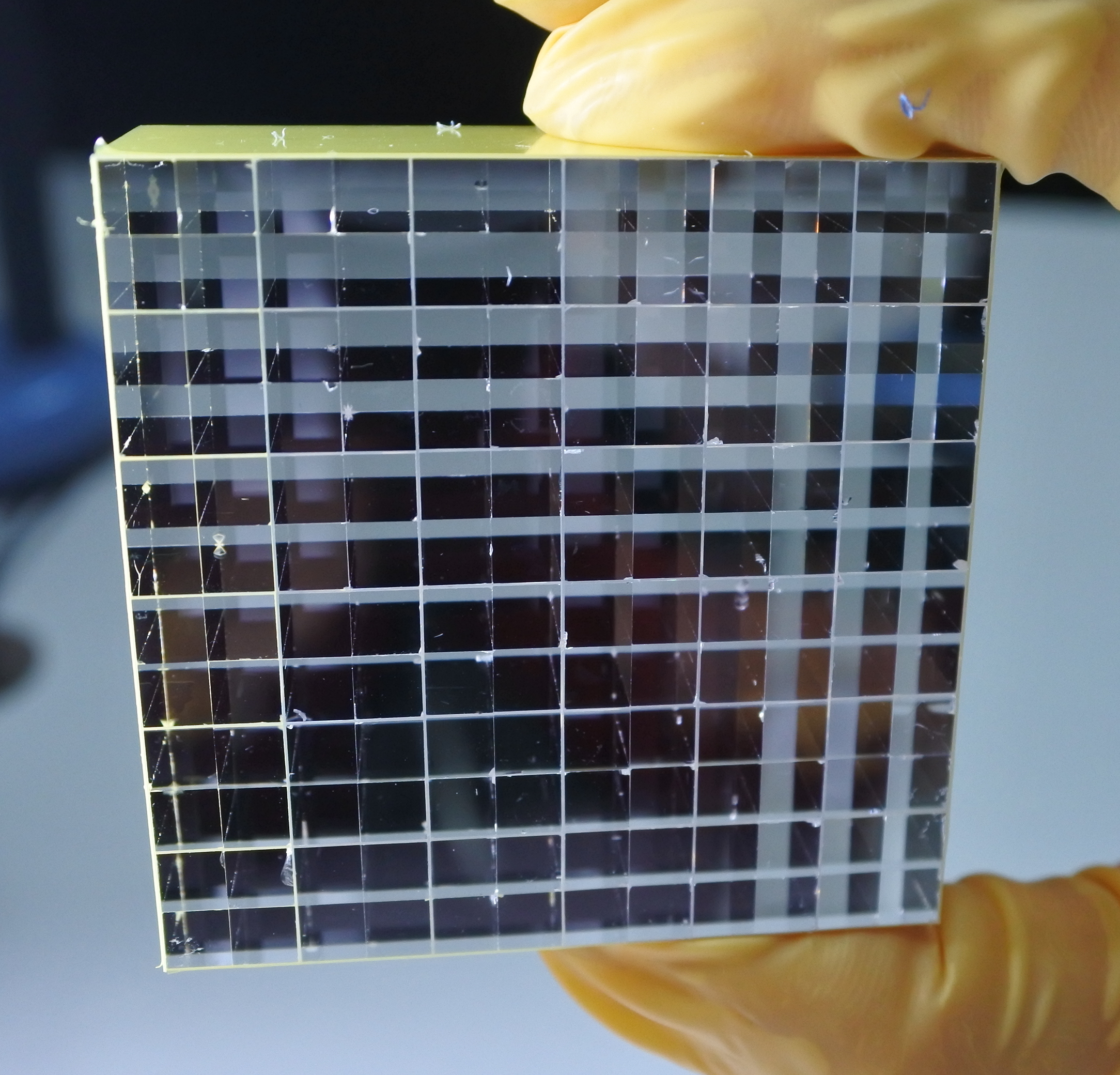}
 	\label{fig:array_assembly}}%
	\hspace{2mm}
	\subfloat[]{\includegraphics[width=0.395\textwidth]{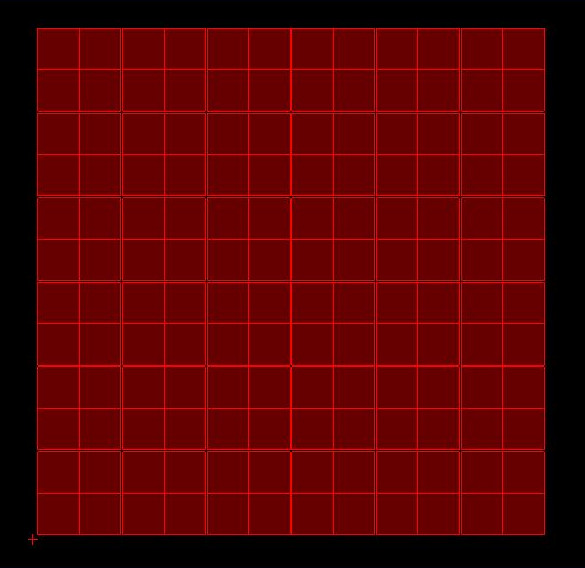}
	\label{fig:simulation_array}}%
	\caption{The crystal array used for calibration is shown in (a) and the corresponding simulation geometry in (b).}
	\label{fig:crystal_array}
\end{figure}

For the validation of the proposed algorithm with a known ground truth, Geant4 simulations of an LSO crystal array with the same geometry as the measurements were created (see Fig. \ref{fig:simulation_array}). This resulted in a crystal array of $6 \times 6$ crystal sub elements, which were separated by a \SI{160}{\micro\meter} gap. Each sub element consisted of four $\SI{3.9}{\milli\meter} \times \SI{3.9}{\milli\meter} \times \SI{16}{\milli\meter}$ LSO crystals. Since no optical simulation was conducted, ESR foil between sub elements and wrapping on the outside of the crystal array were not defined. Energy deposits generated by a 511-keV gamma photon are summed up per crystal element. Here we considered a gamma event to be one $\SI{511}{\kilo\electronvolt}$ gamma interacting with the crystal array. The gamma source was implemented as an isotropically emitting point source centered in front of the crystal array with a distance of \SI{20}{\milli\meter} to illuminate the full array.

\section{Methods}
\label{sec:algorithm}
The proposed algorithm consists of two essential parts. First is a measurement-based calibration, which results in a quantitative knowledge of the light spread for each crystal in the detector in case of a SCI. Secondly, in the recovery step, this knowledge is used to fit a linear combination of light spread functions in order to reproduce the light pattern of a given measurement.
\subsection{Calibration}
\begin{figure}[h!]
	\centering
	\subfloat[]{\includegraphics[width=0.2\textwidth]{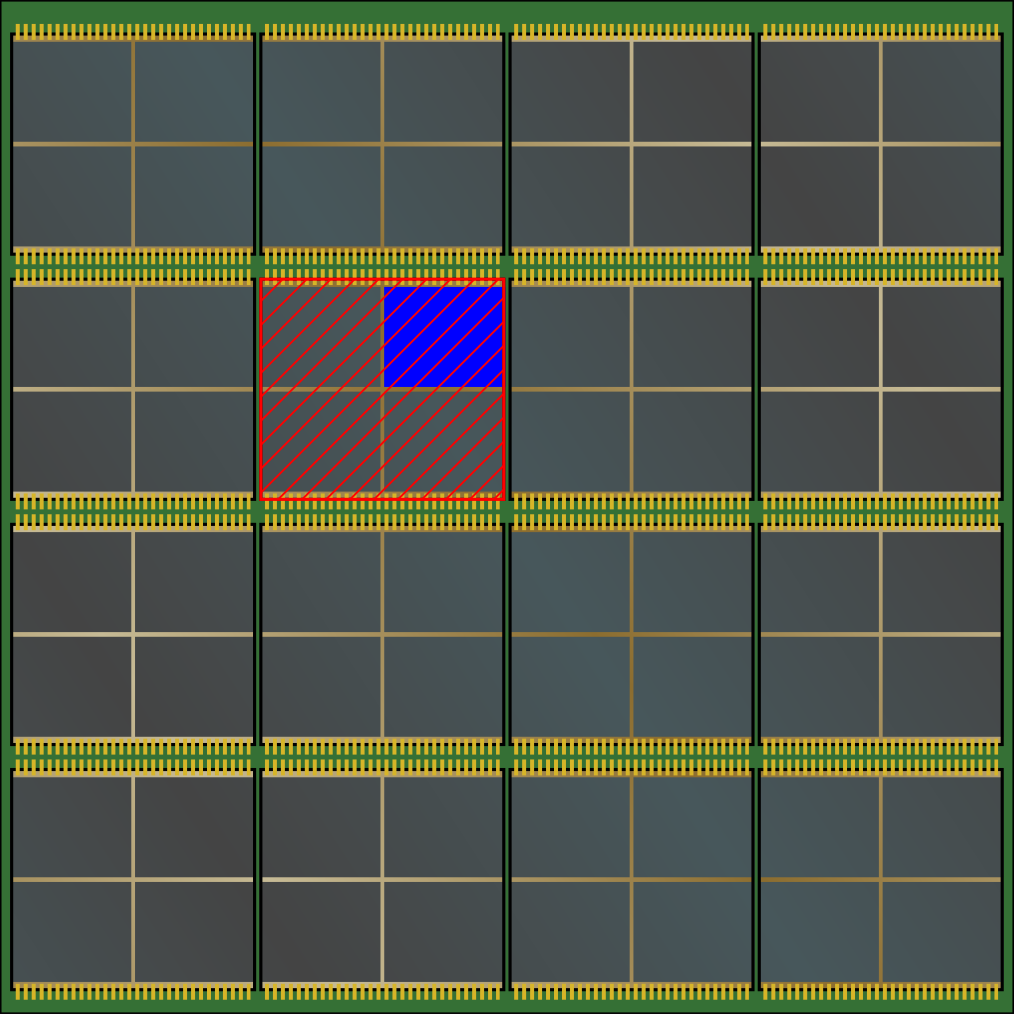}}
	\hspace{20pt}
	\subfloat[]{\includegraphics[width=0.2\textwidth]{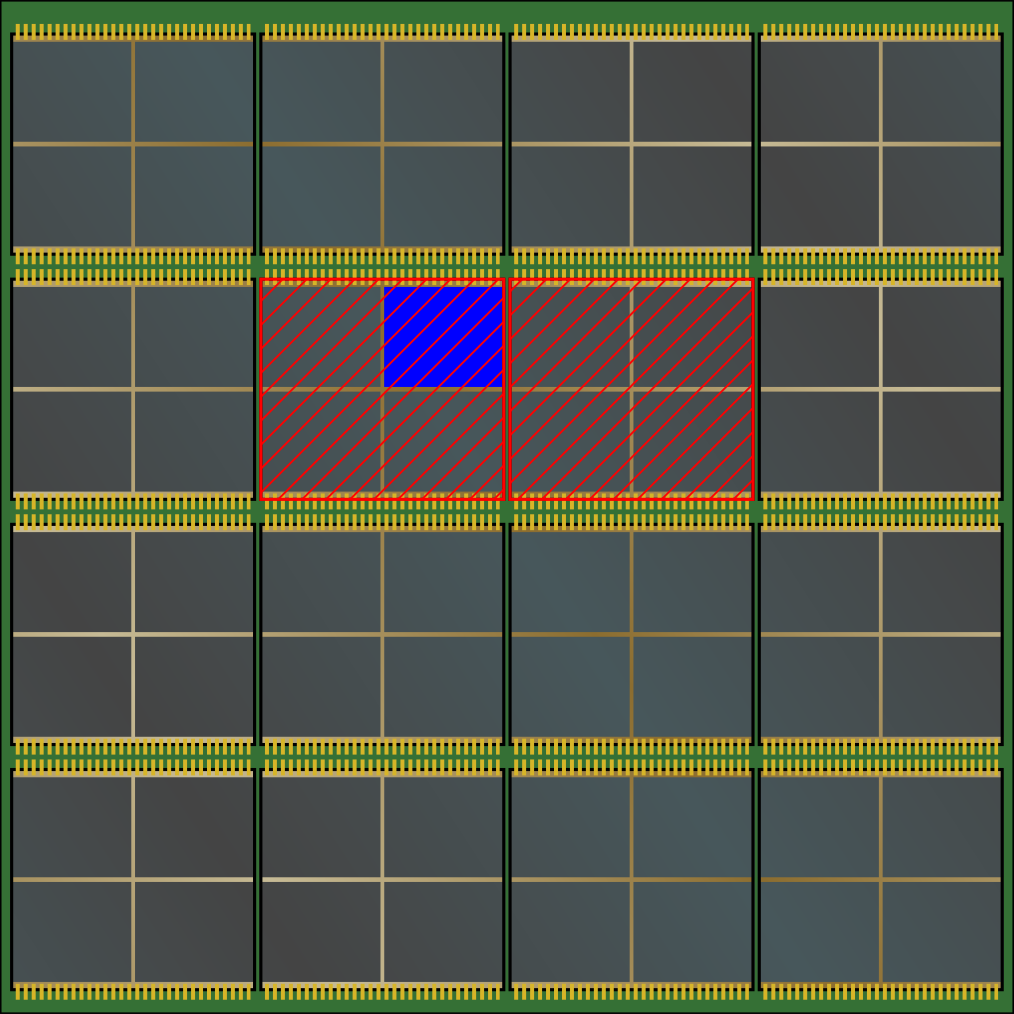}}
	\hspace{20pt}		\subfloat[]{\includegraphics[width=0.2\textwidth]{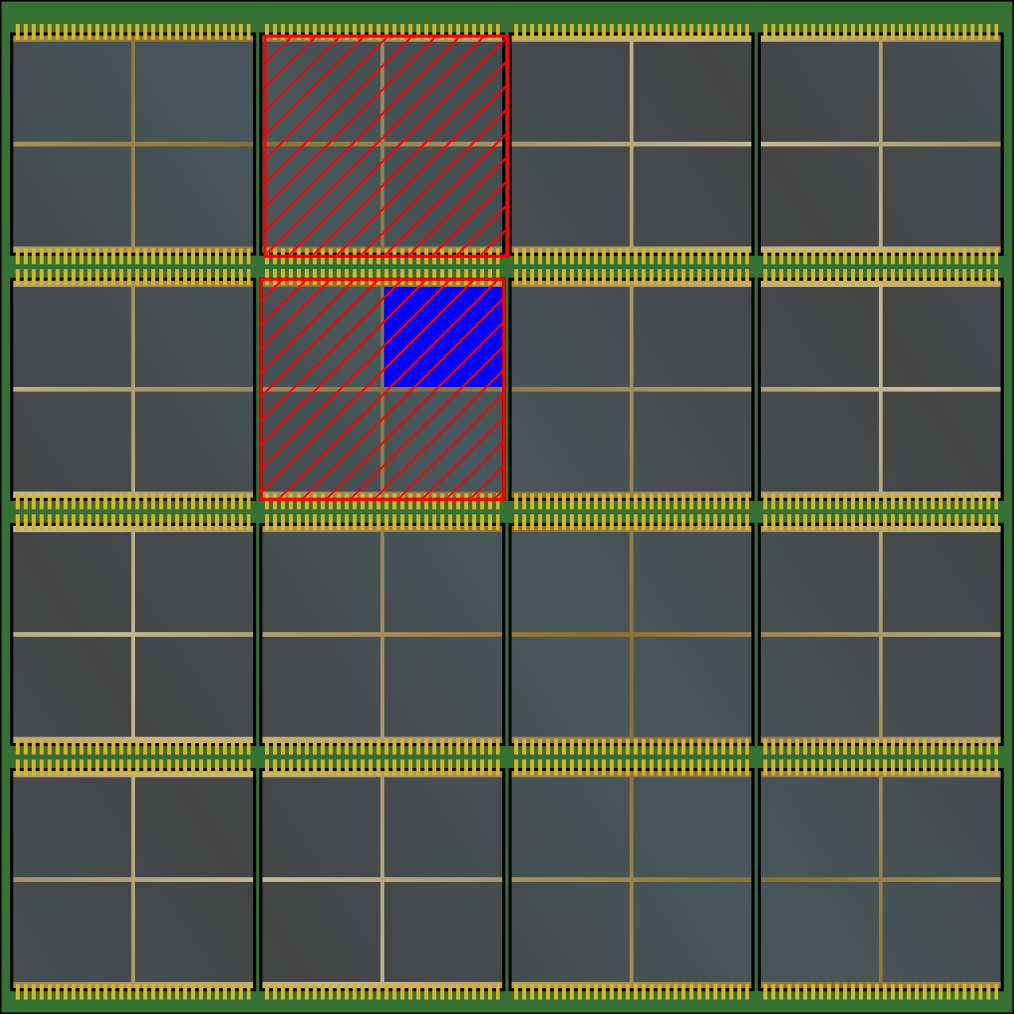}}
	\hspace{20pt}
	\subfloat[]{\includegraphics[width=0.2\textwidth]{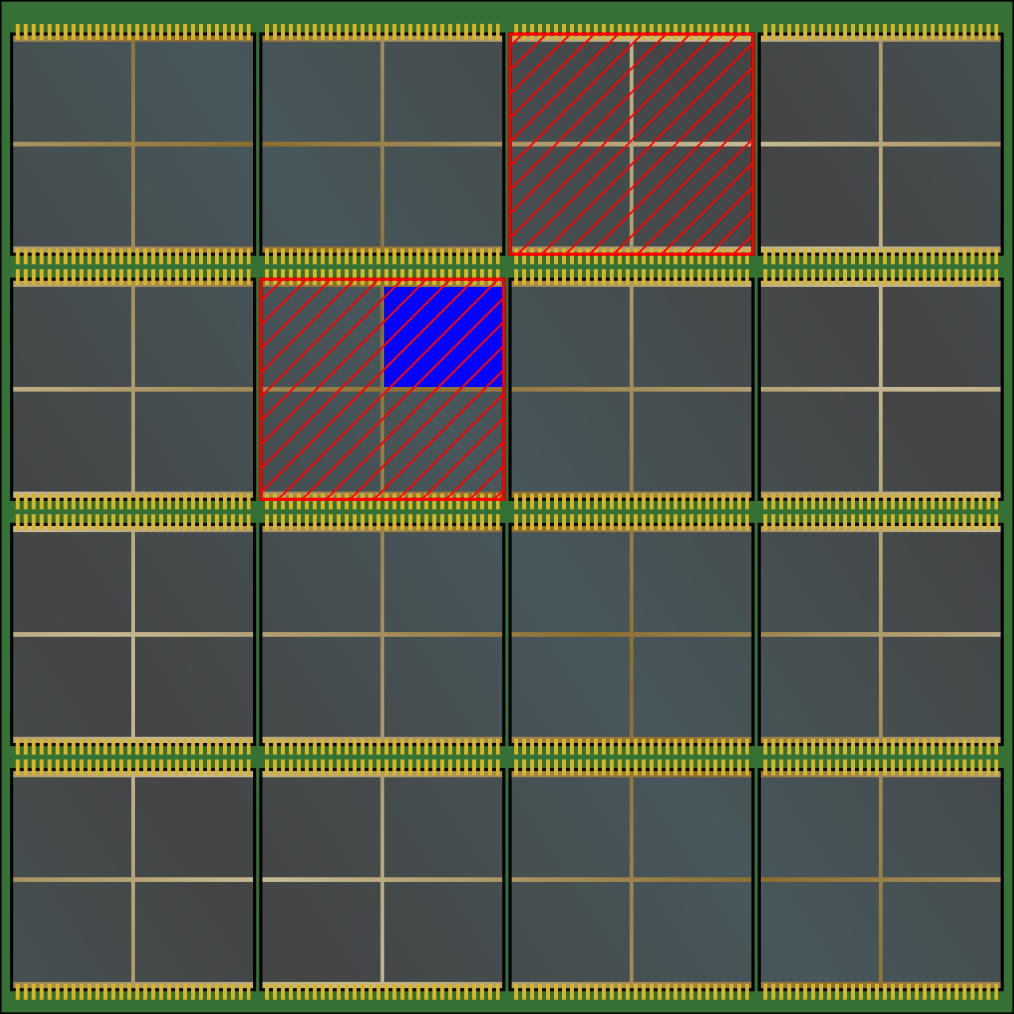}}
	\vfill
	\subfloat[]{\includegraphics[width=0.25\textwidth]{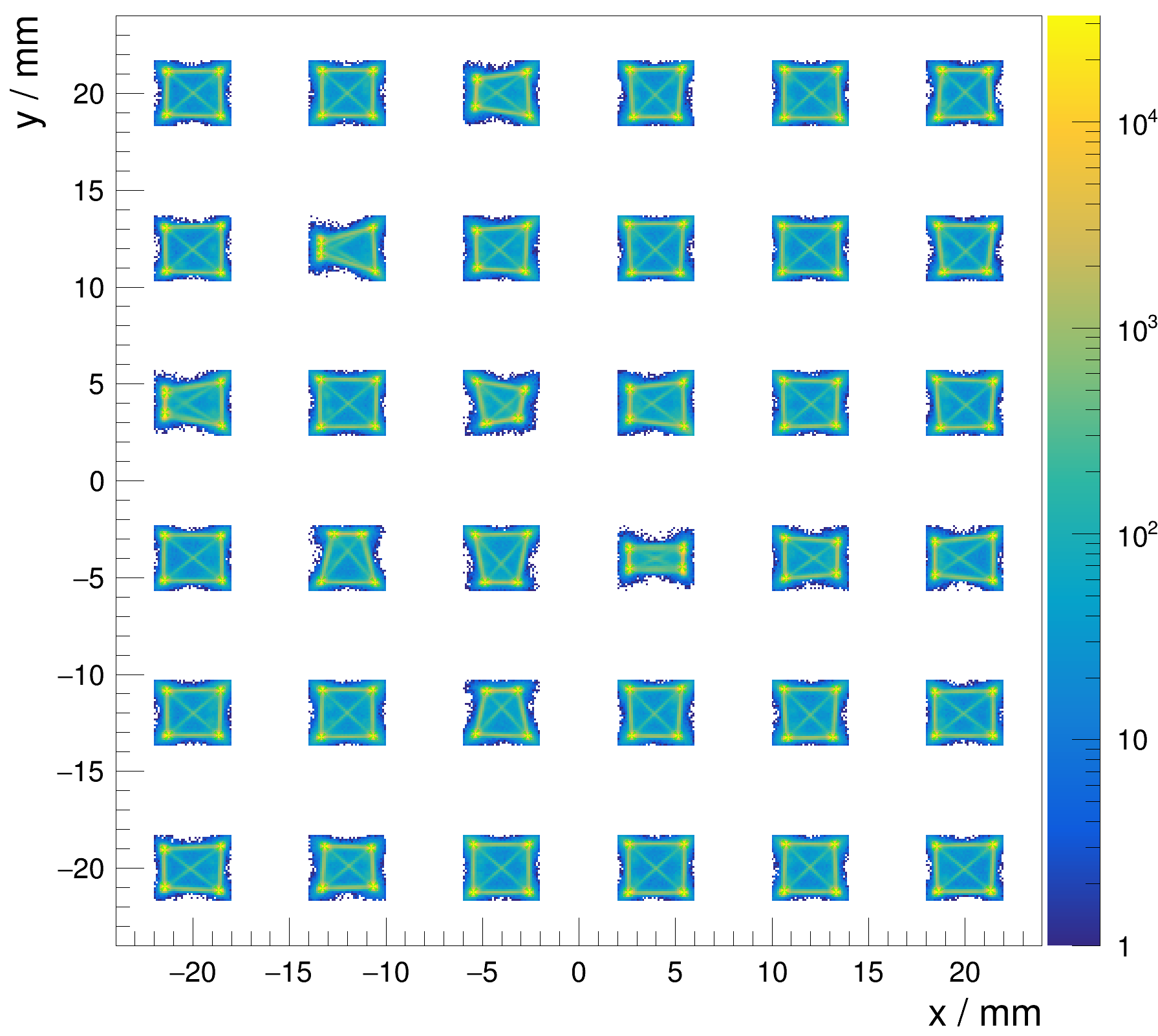}}
	\hfill
	\subfloat[]{\includegraphics[width=0.25\textwidth]{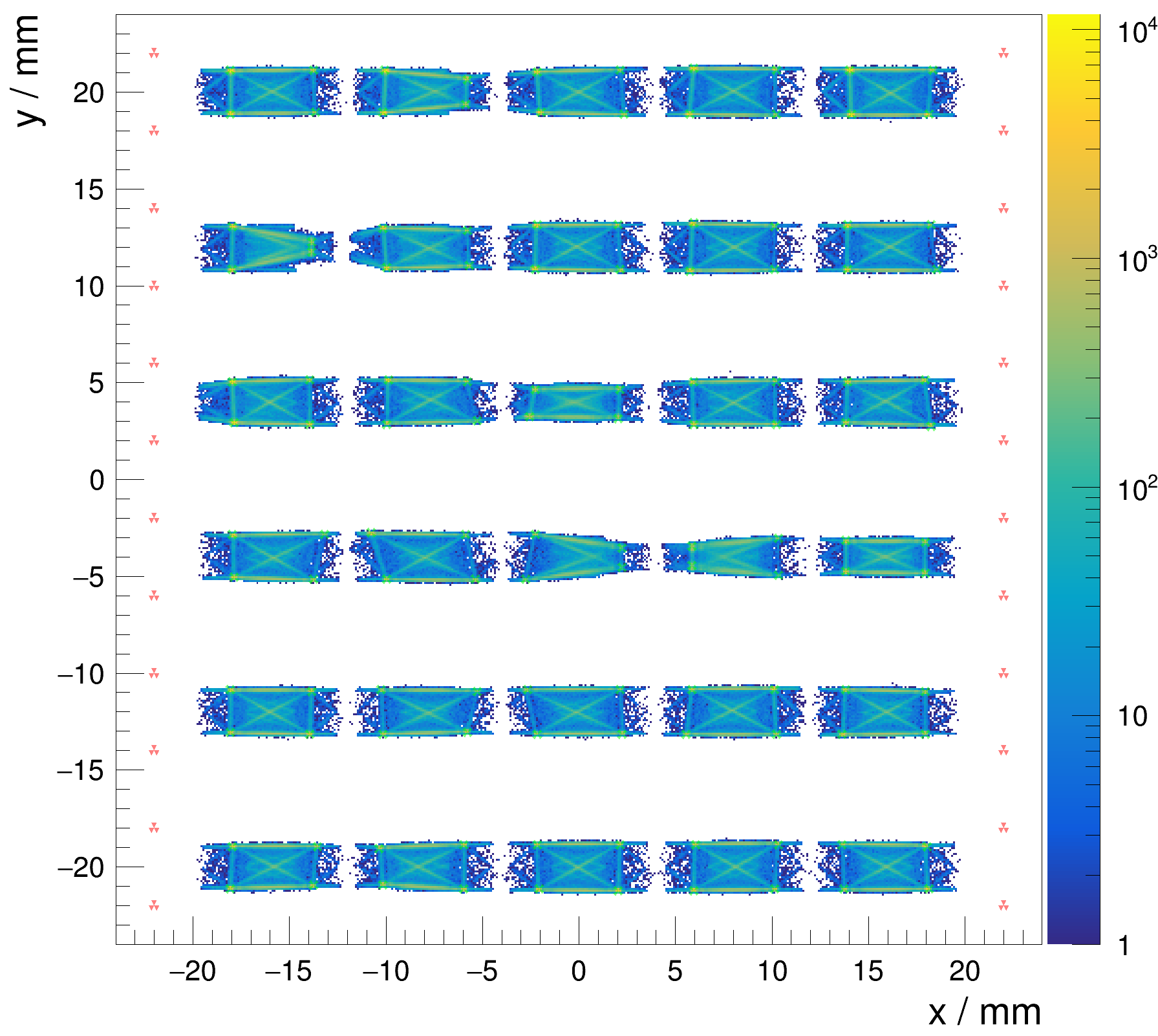}}
	\hfill
	\subfloat[]{\includegraphics[width=0.25\textwidth]{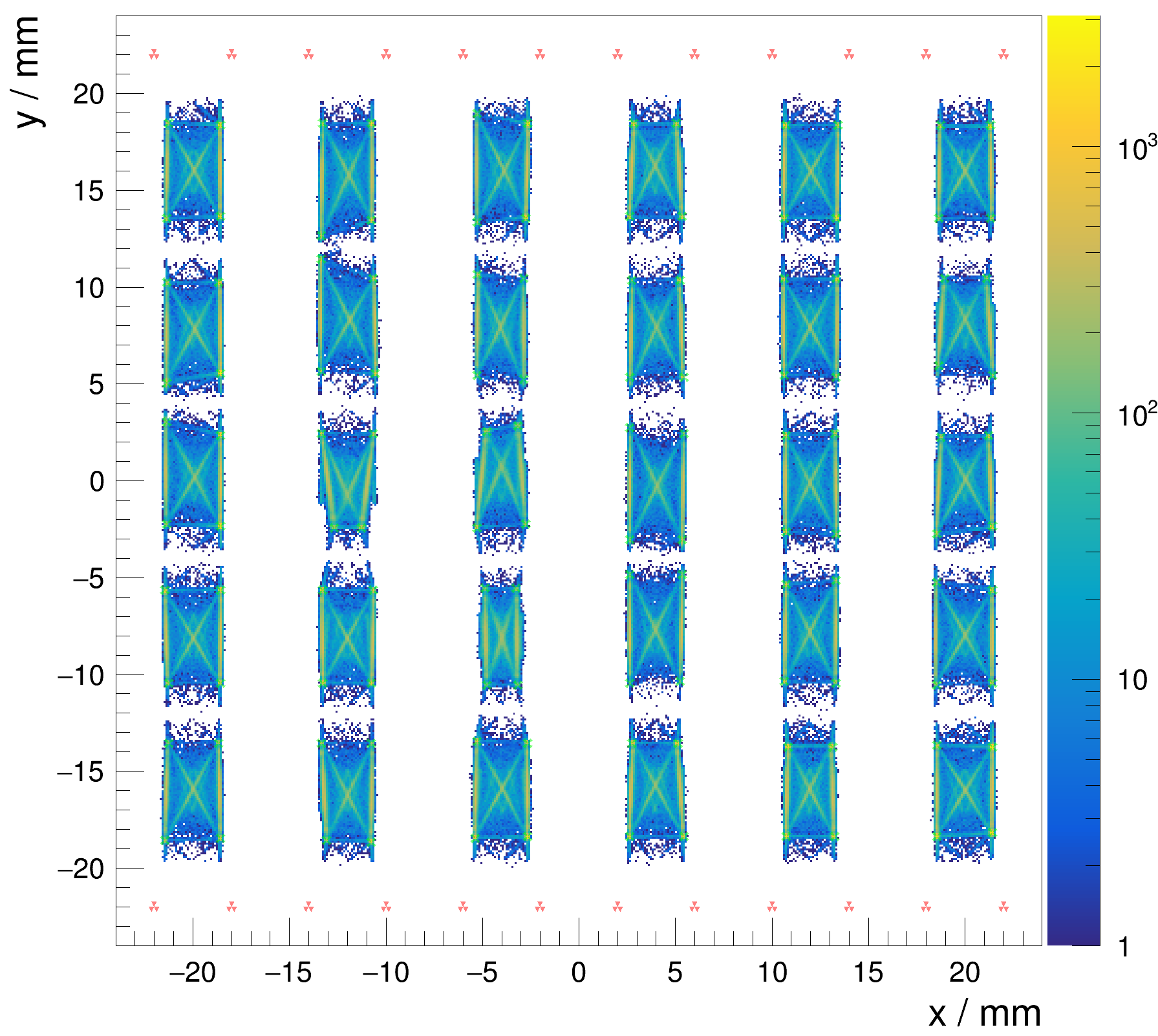}}
	\hfill
	\subfloat[]{\includegraphics[width=0.25\textwidth]{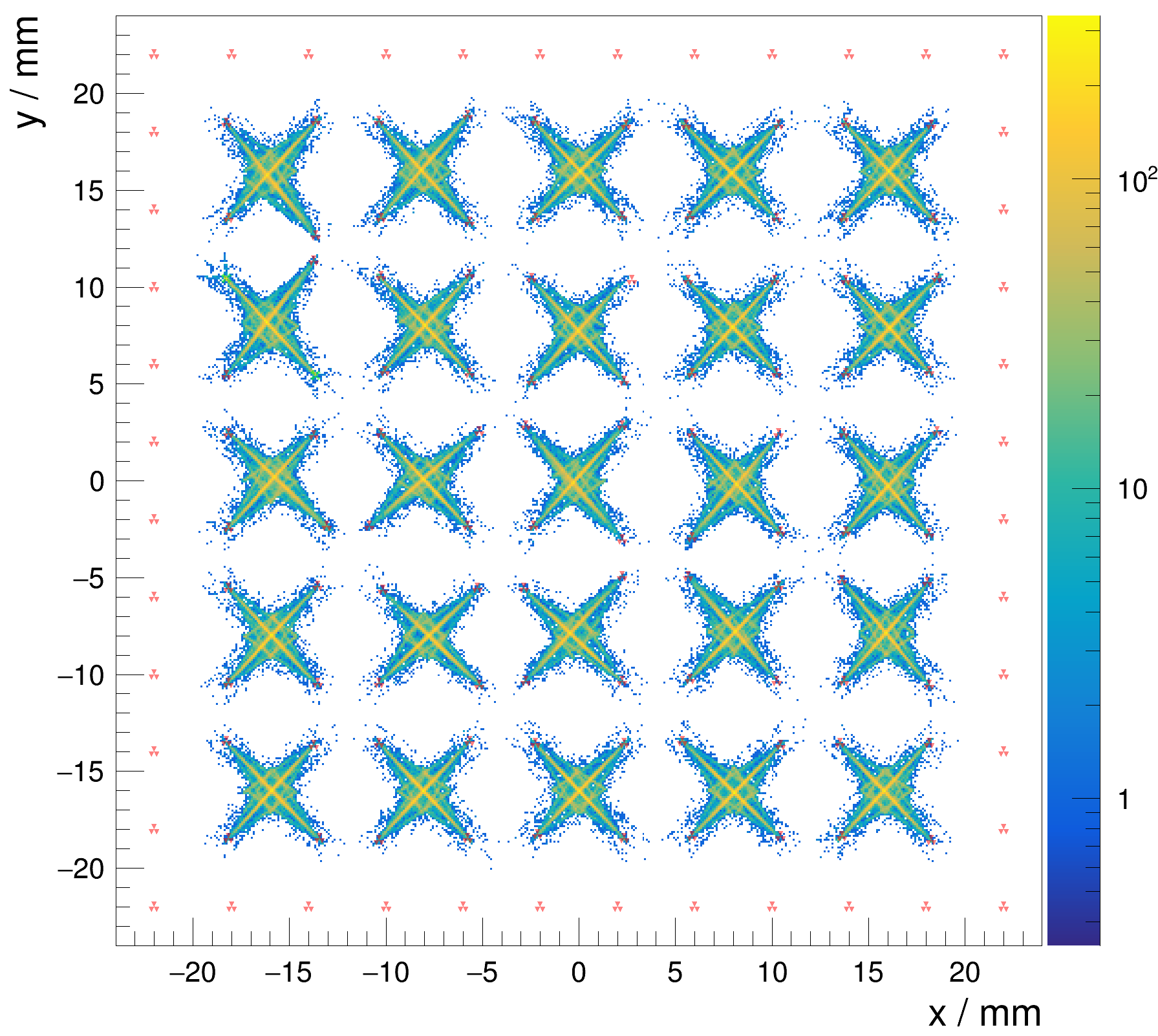}}
	
	\caption{The left column shows an exemplary location of a main pixel marked in blue and the respective ROI for different neighbour combinations specifically for a DPC configuration of $4 \times 4$ DPCs. The right column shows for that COG neighbour ROI the floodmap histogram of all main pixels on the sensor tile. a) and e) depict the main dSiPM criterion, b) and f) the horizontal neighbour criterion, c) and g) the vertical neighbour criterion and d) and h) the diagonal neighbour criterion.}
	\label{fig:roi_floodmaps}
\end{figure}
\begin{figure}[!h]
	\centering
	\subfloat[]{\includegraphics[width=0.4\textwidth]{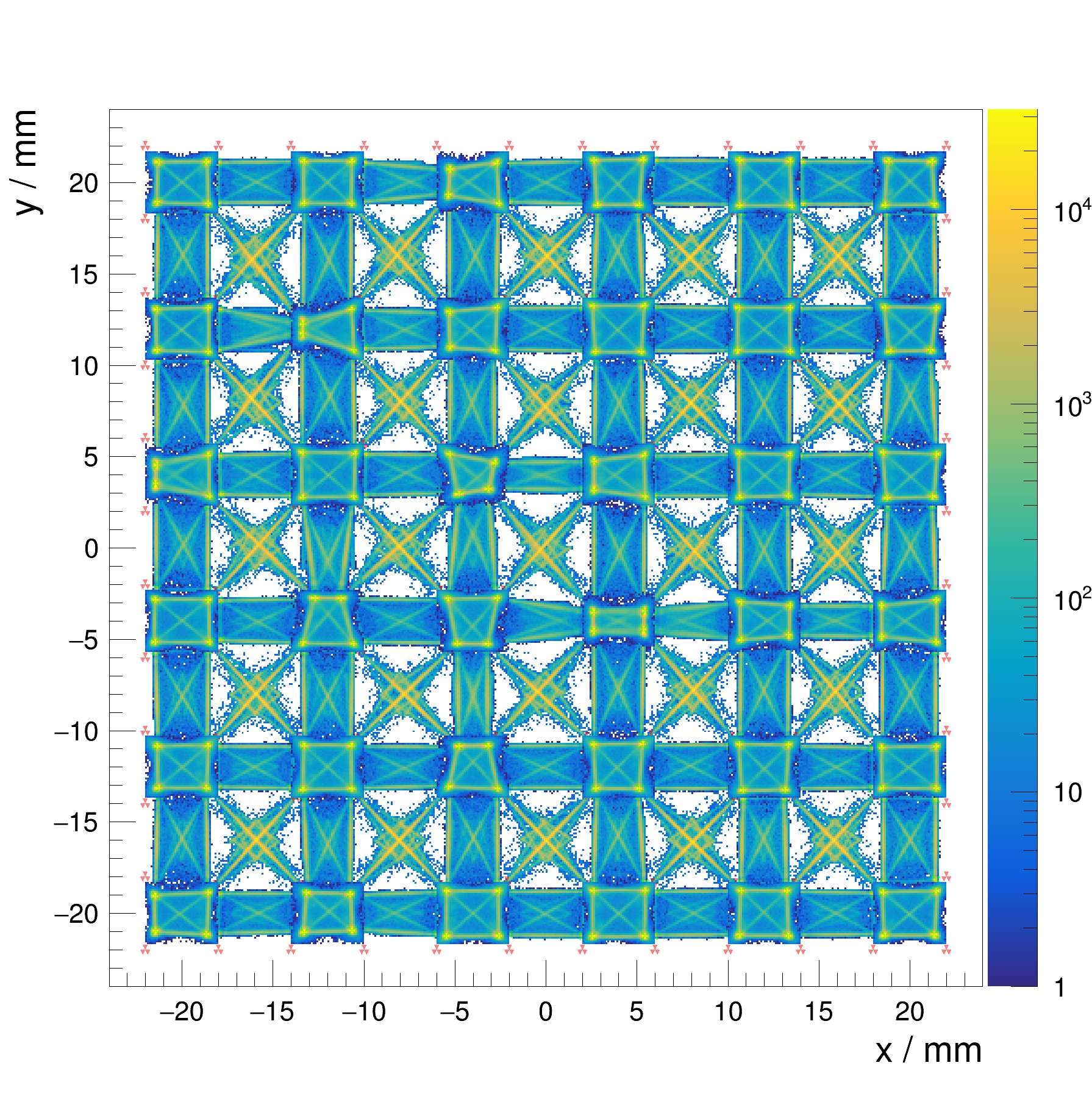}%
		\label{fig:floodmap_unfiltered}}
	\hspace{20pt}
	\subfloat[]{\includegraphics[width=0.4\textwidth]{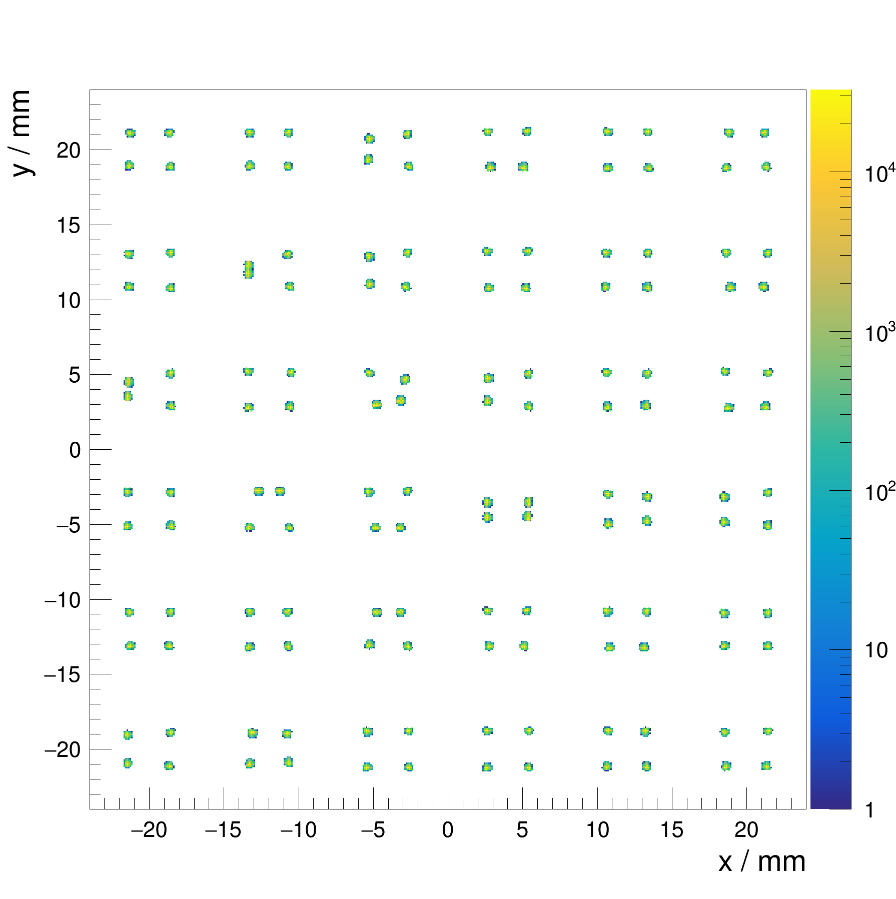}
		\label{fig:floodmap_sci_filter}}
	\caption{In (a), a superposition of flood maps for all different DPC readout conditions is shown. The peaks correspond to the crystal positions and SCIs, while the diagonal crosses represent ICS and light sharing. (b) shows the strictly filtered SCIs at the crystal positions for the main dSiPM criterion. It becomes apparent in both figures that the light distribution varies from the mean behaviour due to differences in the optical coupling of the four crystals located on a single DPC for some crystals (see e.g. crystal at $5$,$-5$), which cannot be achieved with a perfect optical simulation and therefore motivates the measurement-based approach.}
	\label{fig:calibration_floodmaps}
\end{figure}

The calibration starts by analysing flood measurements, which are conducted as described before. The resulting signals are clustered by their timestamps within a singles cluster window of \SI{30}{\nano\second}. With the help of a center of gravity (COG) algorithm the clusters are assigned a position on the detector. Using the location of the pixel with the highest photon count, the main pixel, regions of interest (ROI) are defined. The ROIs are based on the DPCs housing the neighboring pixels to the main pixel. For central main pixel locations there is a horizontal, a vertical and a diagonal neighbor DPC. For main pixels located on the edge or the corner, not all neighbours are physically present. Considering three neighbouring DPCs we can define eight ROIs with different combinations of the neighbouring DPCs. The COG of all pixels of the DPCs belonging to a ROI is then calculated if the respective DPCs are present in the cluster, meaning they have triggered, were validated and their timestamp was within the cluster window. Up to eight different two dimensional COG coordinates can be calculated for each cluster. An example of these COG coordinates can be found in Schug et al (2015) \cite{Schug2015}, where two of the eight possible coordinates were computed. Some exemplary ROIs are shown in Figure \ref{fig:roi_floodmaps}. The COG positions are then filled into eight separate 2D flood map histograms (see Fig. \ref{fig:roi_floodmaps} and Fig. \ref{fig:floodmap_unfiltered}, where SCIs can be seen in the high-count areas and the connecting lines represent ICS events) according to the ROI they are assigned to. Based on the fitted positions, geometrical cuts in the individual COG variables can be defined to generate the SCI-LSM. With the described method we are adaptive to the DPC triggering situation and optimize the rejection efficiency by using the largest possible cluster size defined by the ROIs. With a further rejection of clusters with DPCs triggered outside the ROIs we remove ICS events with a large distance between the crystals that generate scintillation light and reject pile up events as well (see Fig. \ref{fig:floodmap_sci_filter} as an example of one filtered flood map). Figure \ref{fig:roi_floodmaps} and Figure \ref{fig:calibration_floodmaps} both show that there is an inherent variation in the light distribution (see e.g. crystal at $5$,$-5$), where the COG peaks are closer together and asymmetrical in comparison to the rest of the array. This makes the importance of an individual, measurement-based calibration for each detector apparent. These effects cannot be easily modelled with an optical simulation, which typically assumes a regular and perfect optical behaviour over the whole crystal-SiPM detector block.\\
Using the filtered SCI events and exploiting the one-to-one coupling scheme, we assign the crystal ID based on the main pixel ID. For each crystal, we histogram the photon sum of the four pixels of the main DPC, the value used for energy estimation, into one spectrum and the light fraction captured by each readout channel into individual histograms. Normalization is performed by dividing the readout channel's photon count by the value used for energy estimation. A timestamp for the cluster is set to the timestamp of the main DPC.\\
The SCI-LSM consists of the means of the light fraction histograms, while an uncertainty matrix represents the width of these histograms. In case of the used detector with one-to-one coupled crystals this is a $144 \times 144$ matrix for the light fractions and a matrix of the same size for the uncertainties as well as a vector of 144 photon-to-energy conversion factors.
\subsection{Recovery}
\label{subsubsec:recovery}
As a first step the LSM is reduced to the triggered channels and crystals matching these channel IDs. Only crystals whose calibrated light patterns include the triggered photosensors are considered to have contributed to the measured light pattern.\\
A numerical least-square method is used to solve the matrix equation Eq. \ref{eq:matrix_equation}, where $\boldsymbol{\mathit{M}}$ describes the LSM, $\vec{e}$ is the energy distribution vector of the crystals and $\vec{p}$ is the measured light distribution, and find the most-probable solution for $\vec{e}$ to describe the observed photon pattern $\vec{p}$.
\begin{equation}
	\label{eq:matrix_equation}
	\boldsymbol{\mathit{M}} \cdot \vec{e} = \vec{p}
\end{equation}
As the solver is unconstrained, negative energy values can be part of the solution. Therefore, after each solving of Eq. \ref{eq:matrix_equation}, the contributing crystals in the energy distribution vector are either reduced by removing the lowest negative energy entry or by removing all energy entries below a certain threshold larger than zero, which has been varied. Reducing the number of crystals contributing to the energy distribution vector, requires the size of the LSM to also be reduced. This method is applied iteratively allowing the solution to converge to a stable point with physically sensible energy contributions of a set of crystals over several fitting iterations.\\
We considered two methods to improve algorithm performance. On the one hand, we used weights on the calibration matrix and measurement. The weights tested were based on the number of photons detected per SiPM (photon weights) 
and the standard deviation of the calibrated light fraction distribution per channel (sigma weights)
. On the other hand, we applied an energy filter (posteriori filter), which dismisses all crystals in the algorithm solution with energies below a certain threshold. If this posteriori filter is employed, it affects both the end solution of the algorithm and the metric calculation.

\subsection{Validation}
\label{subsec:validation}
\begin{figure}[h]
	\centering
	\includegraphics[width=0.6\textwidth]{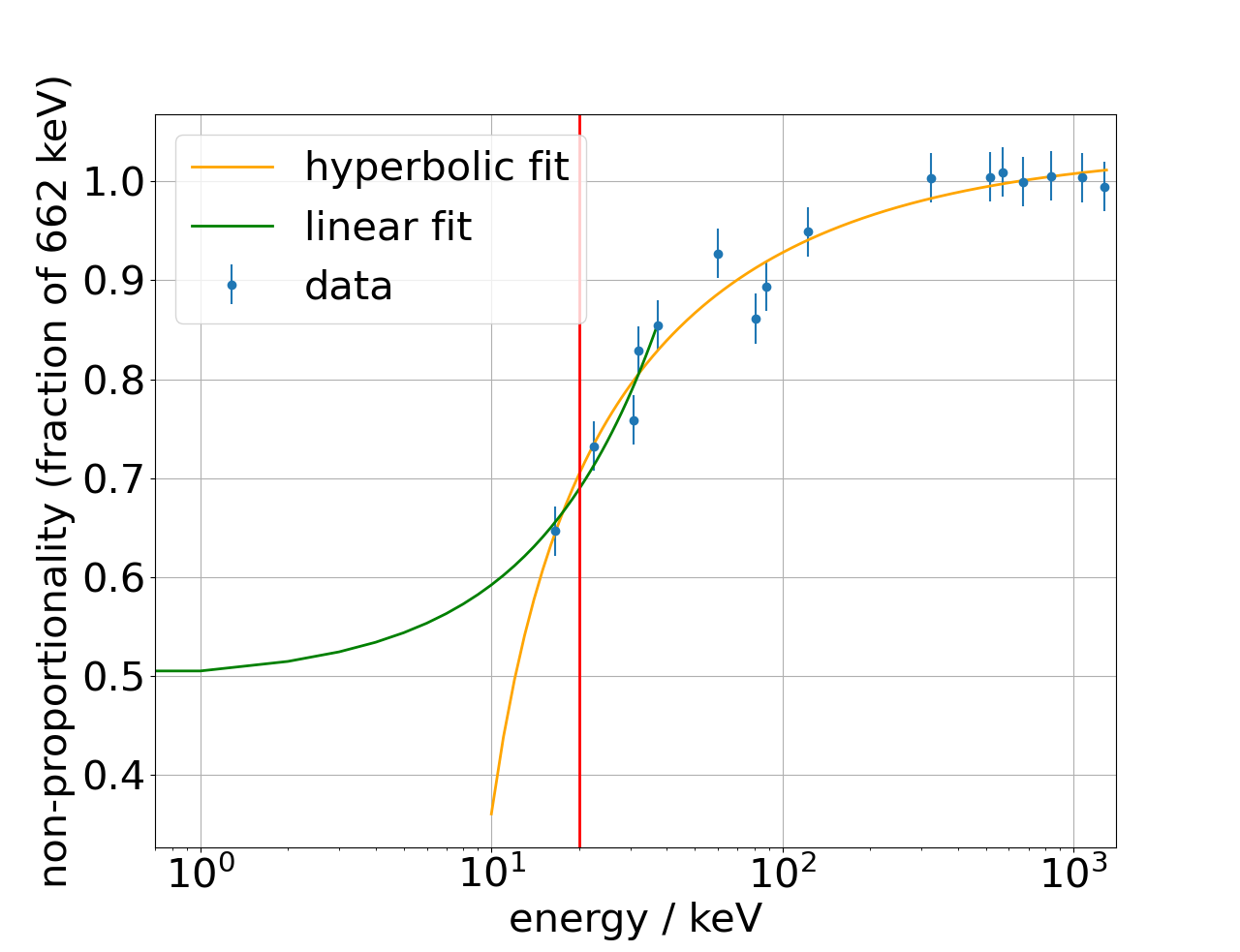}
	\caption{Light yield non-proportionality according to Moszynski et al \cite{Moszynski2016}, down to energies of \SI{16}{\kilo\electronvolt} this follows a hyperbolic function (orange fit). To determine the non-proportionality for energies lower than \SI{16}{\kilo\electronvolt} a linear behaviour is assumed (green fit). In application, the switch from hyperbolic to linear model is made at \SI{20}{\kilo\electronvolt} (red line).}
	\label{fig:non-proportionality_model}
\end{figure}

For the validation of the algorithm's functionality, the simulation output is used as a ground truth. The energy deposits of the simulation are combined with the light spread as described by the calibration matrix in order to model the channel response of the photo-sensor (see Fig. \ref{fig:simulation+lightspread}). Since a numerical method is used to achieve the solution, some deviations are expected. The performance of the algorithm is qualitatively assessed with two metrics:
\begin{itemize}
	\item the first metric ($\Delta_{\textnormal{crystal}}$), assesses the relative deviation from the simulation event energy for each crystal within the algorithm solution or the ground truth:
	\begin{equation}
		\Delta_{\textnormal{crystal}} = \frac{\sum\limits_{i=1}^{N_{\textnormal{crystal}}} { | e^{i}_\textnormal{algorithm} - e^{i}_\textnormal{simulation} |}}{\sum\limits_{i=1}^{N_{\textnormal{crystal}}}{ e^{i}_\textnormal{simulation}}}
	\end{equation}
	\item the second metric ($\Delta_{\textnormal{sum}}$), determines the relative energy sum deviation between the algorithm solution and the simulation:
	\begin{equation}
	\Delta_{\textnormal{sum}} = \frac{\sum\limits_{i=1}^{N_{\textnormal{crystal}}} {e^{i}_\textnormal{algorithm}} - \sum\limits_{i=1}^{N_{\textnormal{crystal}}} {e^{i}_\textnormal{simulation}} }{\sum\limits_{i=1}^{N_{\textnormal{crystal}}}{ e^{i}_\textnormal{simulation}}}
	\end{equation}
\end{itemize}
For the best performance of the algorithm both metrics should be as close to zero as possible.\\
Additionally, we determine the fraction of events (correct crystal fraction), for which the algorithm has assigned all the crystals correctly without considering their energy deposition.\\
In order to achieve a more realistic simulation result, uncertainties were included before applying the algorithm. The two modelled effects are the energy resolution of the crystal and its light yield non-proportionality.\\
To model the intrinsic energy resolution of LSO crystals a Gaussian centered around each energy value with a full width at half maximum (FWHM) of \SI{8}{\percent} was sampled randomly adding an uncertainty to the simulation.\\
The light yield non-proportionality of LSO is well described in Moszynski et al (2016) \cite{Moszynski2016}. According to their non-proportionality fraction at \SI{662}{\kilo\electronvolt}, we developed a continuous model describing the non-proportionality for all energies (see Fig. \ref{fig:non-proportionality_model}). The model consists of a hyperbolic fit to the data of Moszynski et al (2016) \cite{Moszynski2016}, continued for energies lower than \SI{20}{\kilo\electronvolt} with a linear function. This model is applied to the simulation after it was converted into the channel response of the photo-sensor.

\section{Results}

\begin{figure*}[t]
	\begin{minipage}[c]{0.35\textwidth}
		\centering
		\subfloat[]{\includegraphics[scale=0.355]{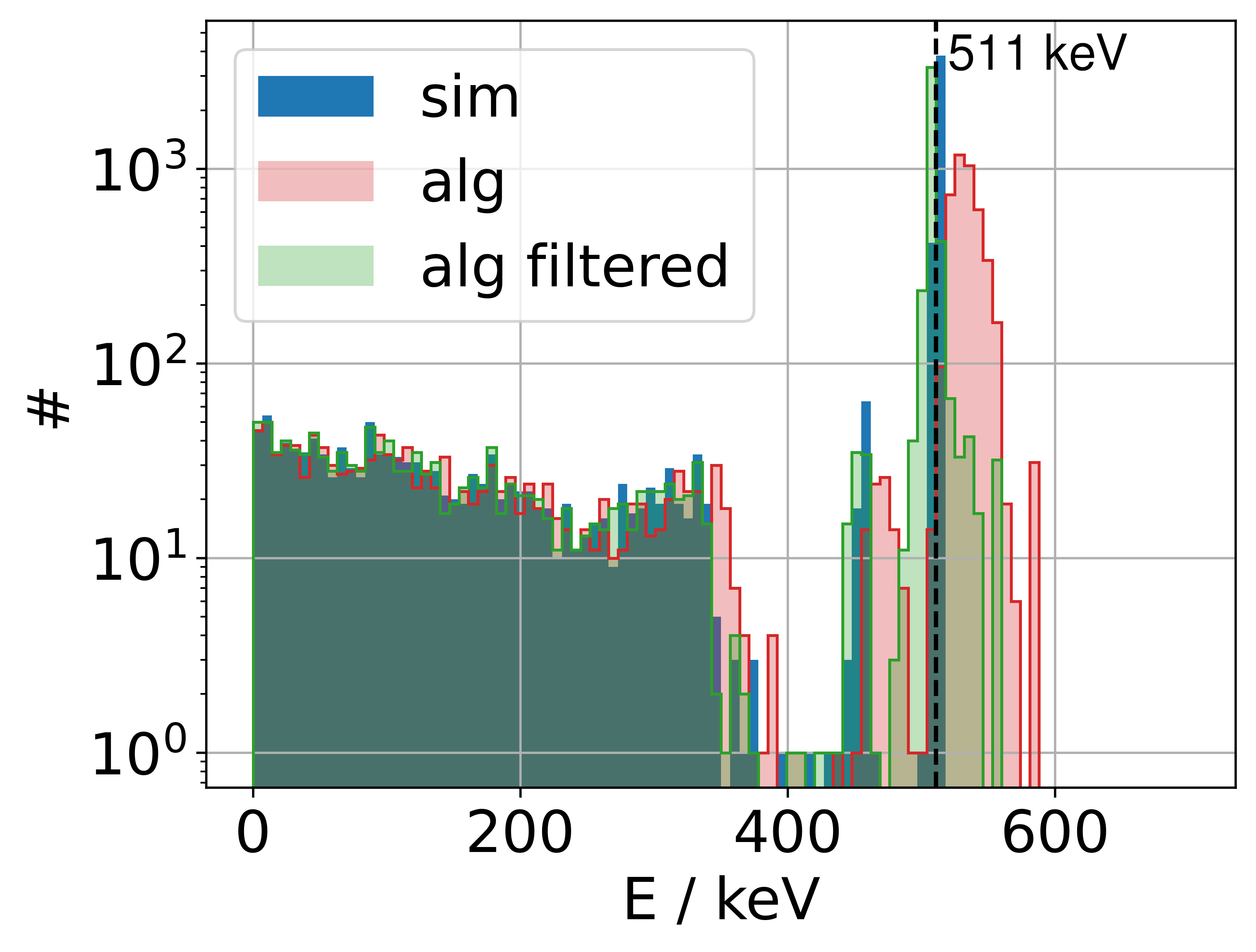}%
		\label{fig:energy_spectra}}
	\end{minipage}%
	\hspace{0mm}
	\begin{minipage}[c]{0.61\textwidth}
		\centering
			\subfloat[]{\includegraphics[scale=0.3]{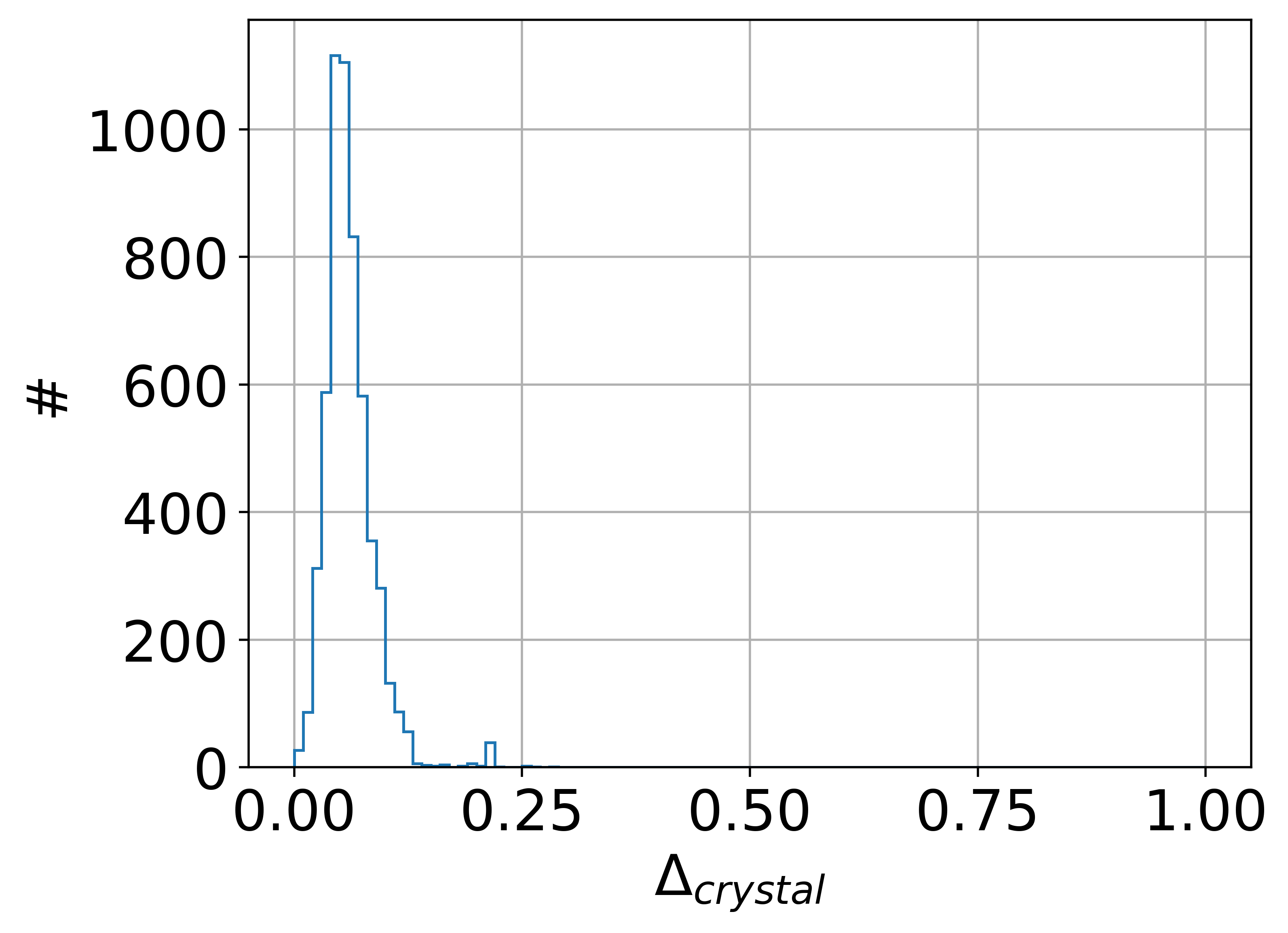}
		\label{fig:metric1_unfiltered}}
	\subfloat[]{\includegraphics[scale=0.3]{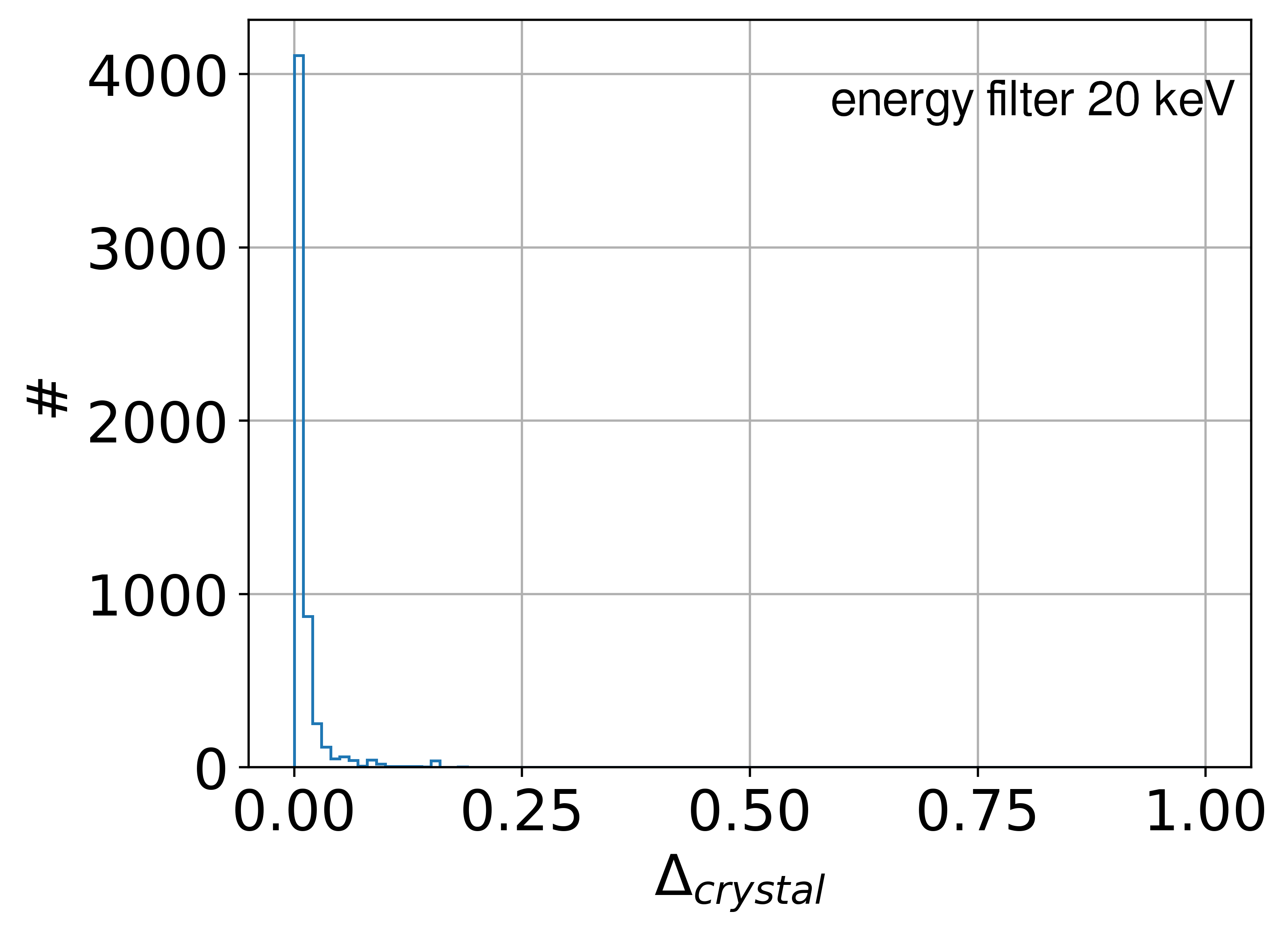}
		\label{fig:metric1_filtered20keV}}
	\vfill
	\subfloat[]{\includegraphics[scale=0.3]{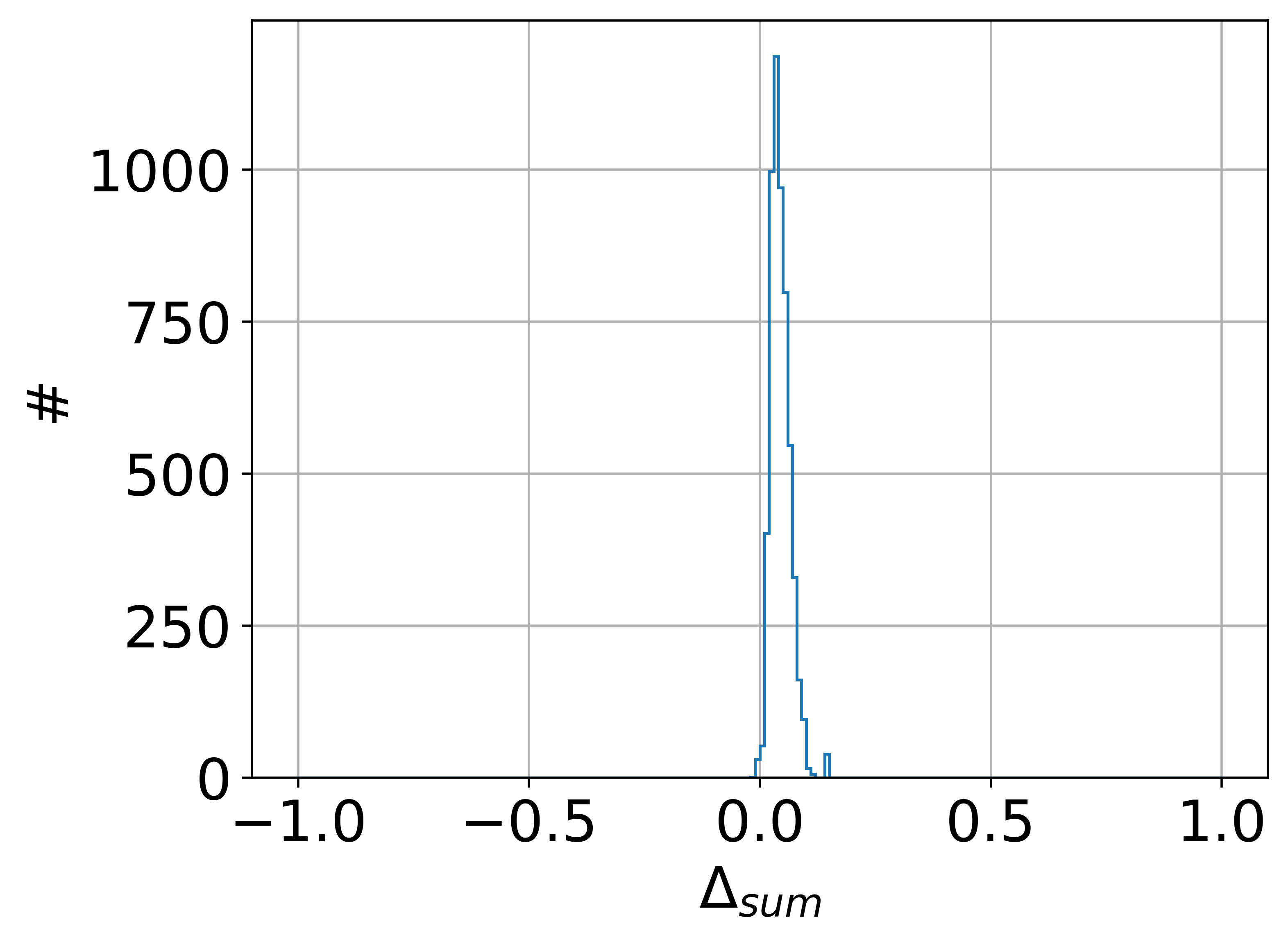}
		\label{fig:metric2_unfiltered}}
	\subfloat[]{\includegraphics[scale=0.3]{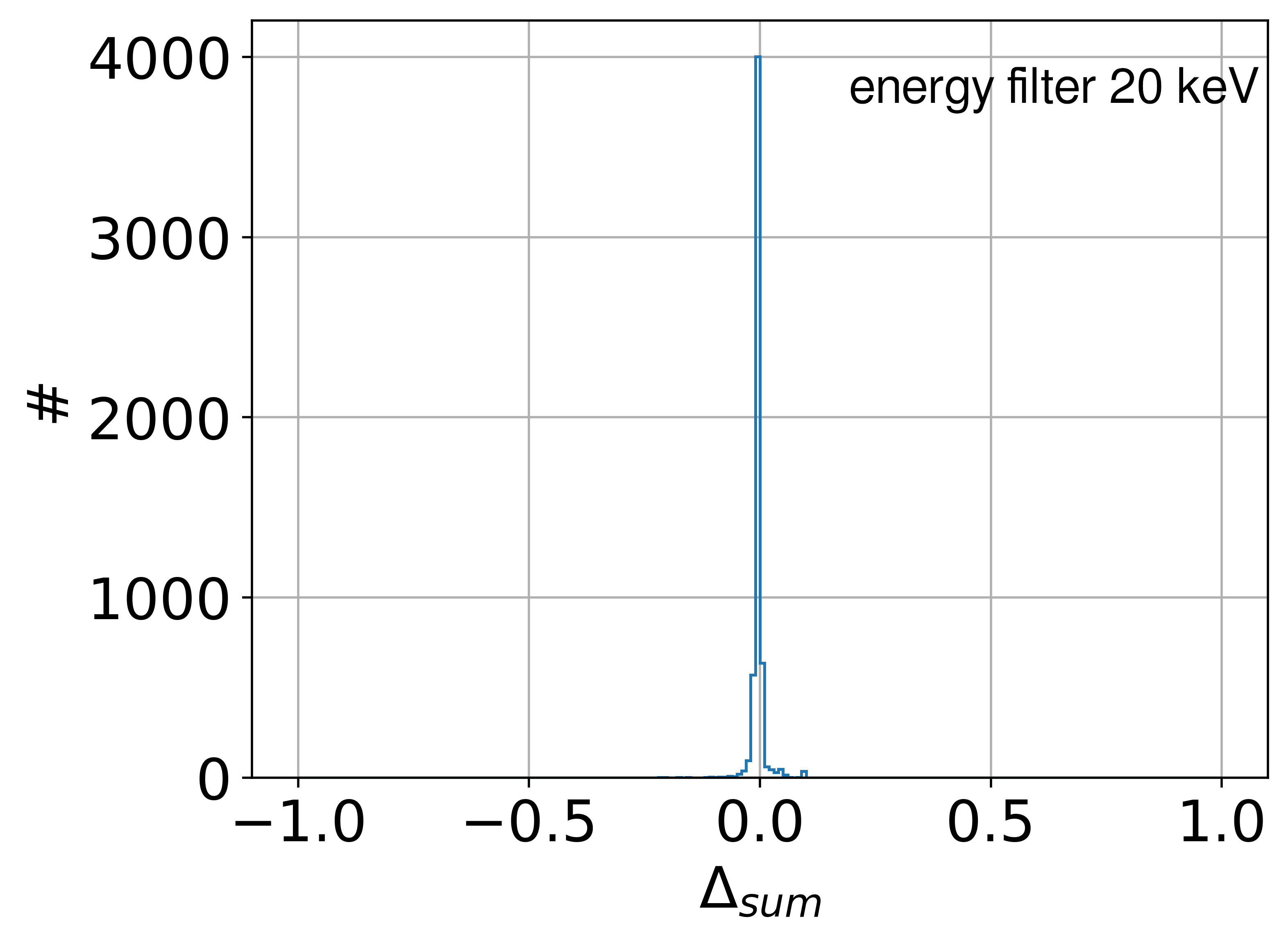}
		\label{fig:metric2_filtered20keV}}
	\end{minipage}%
	\caption{In (a) the event energy sum spectra of the simulation, the algorithm solution and the algorithm solution excluding crystal contributions below $\SI{20}{\kilo\electronvolt}$ are shown. The algorithm solution spectrum shows a structure similar to the simulation, but consistently overestimates energy, which can be seen by the shift of the peaks to higher energies. If crystal contributions with low energies are removed from the energy sum, this effect is significantly reduced. (b) and (d) demonstrate the distribution of the two metrics without energy filter. In (c) and (e), on the other hand, the metric distributions with $\SI{20}{\kilo\electronvolt}$ energy filter can be seen. The filter shifts the distributions closer to zero and sharpens them significantly.}
	\label{fig:results1}
\end{figure*}
\subsection{Basic simulation}
\begin{figure}[h]
	\centering
	\subfloat[]{\includegraphics[width=0.3\textwidth]{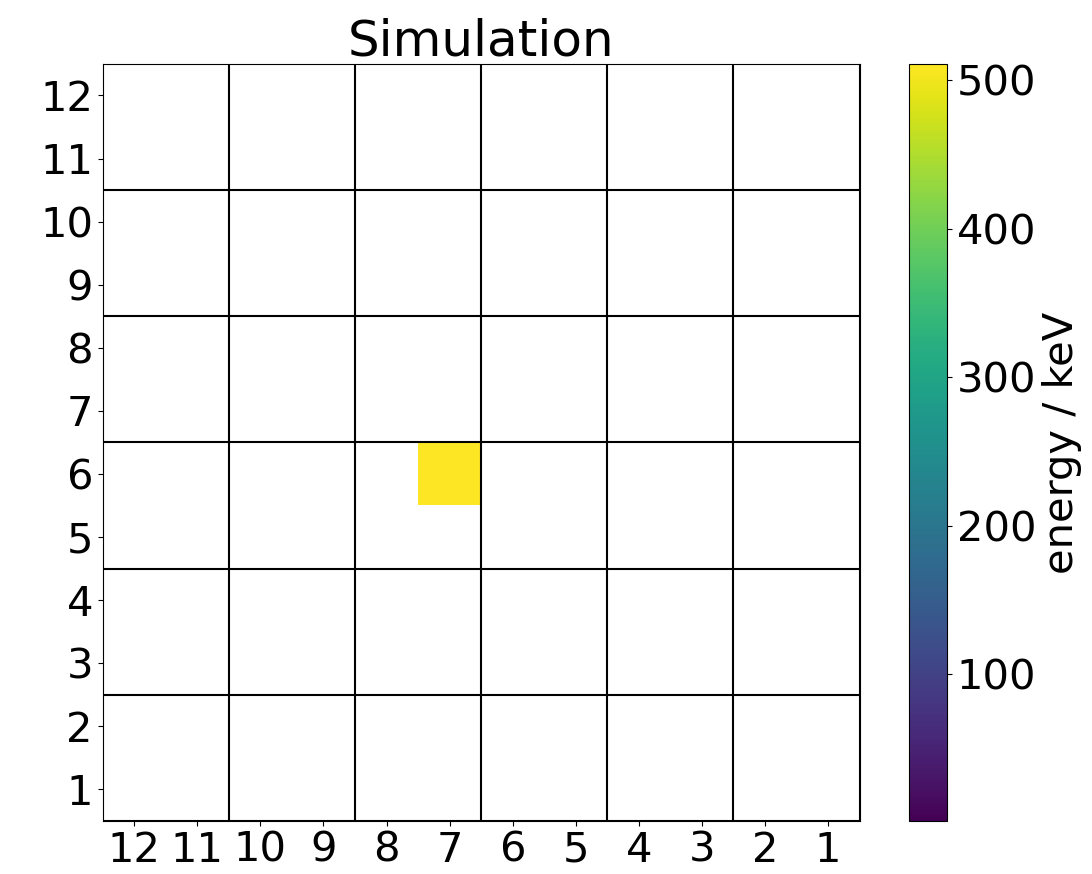}
		\label{fig:simulation}}
	\subfloat[]{\includegraphics[width=0.3\textwidth]{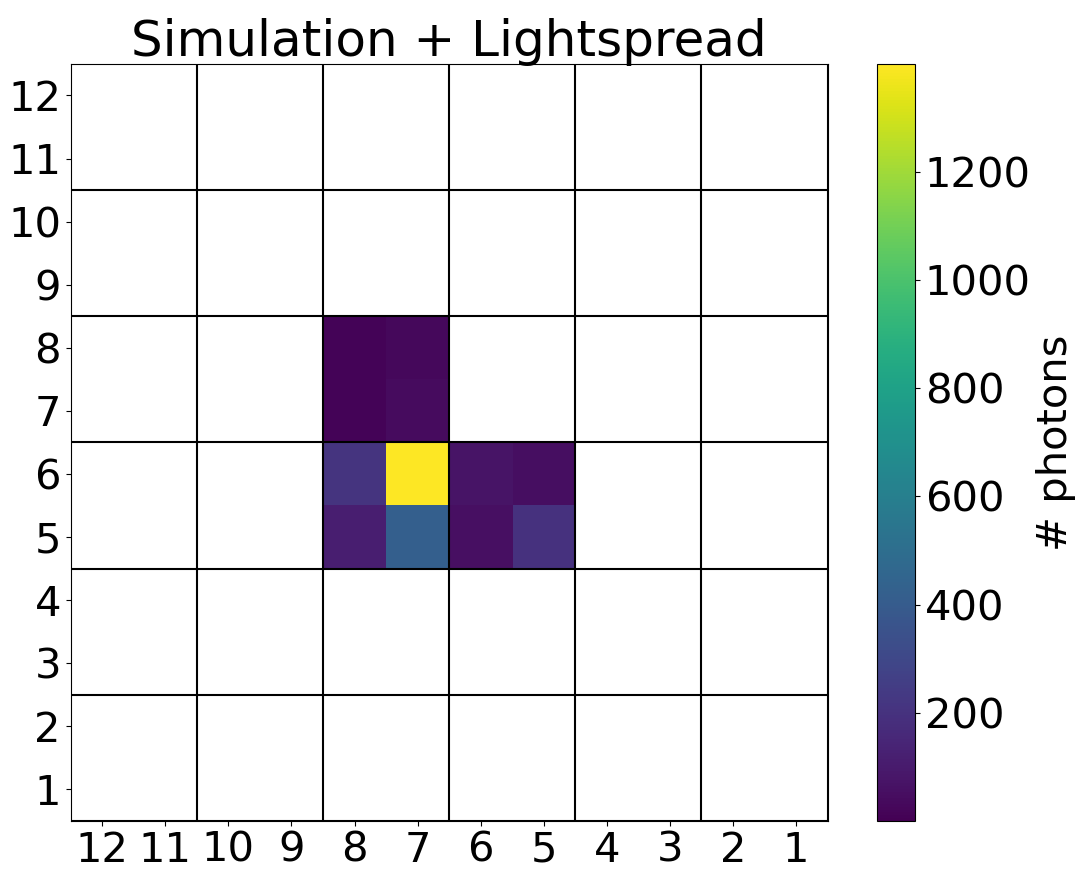}%
		\label{fig:simulation+lightspread}}
	\subfloat[]{\includegraphics[width=0.3\textwidth]{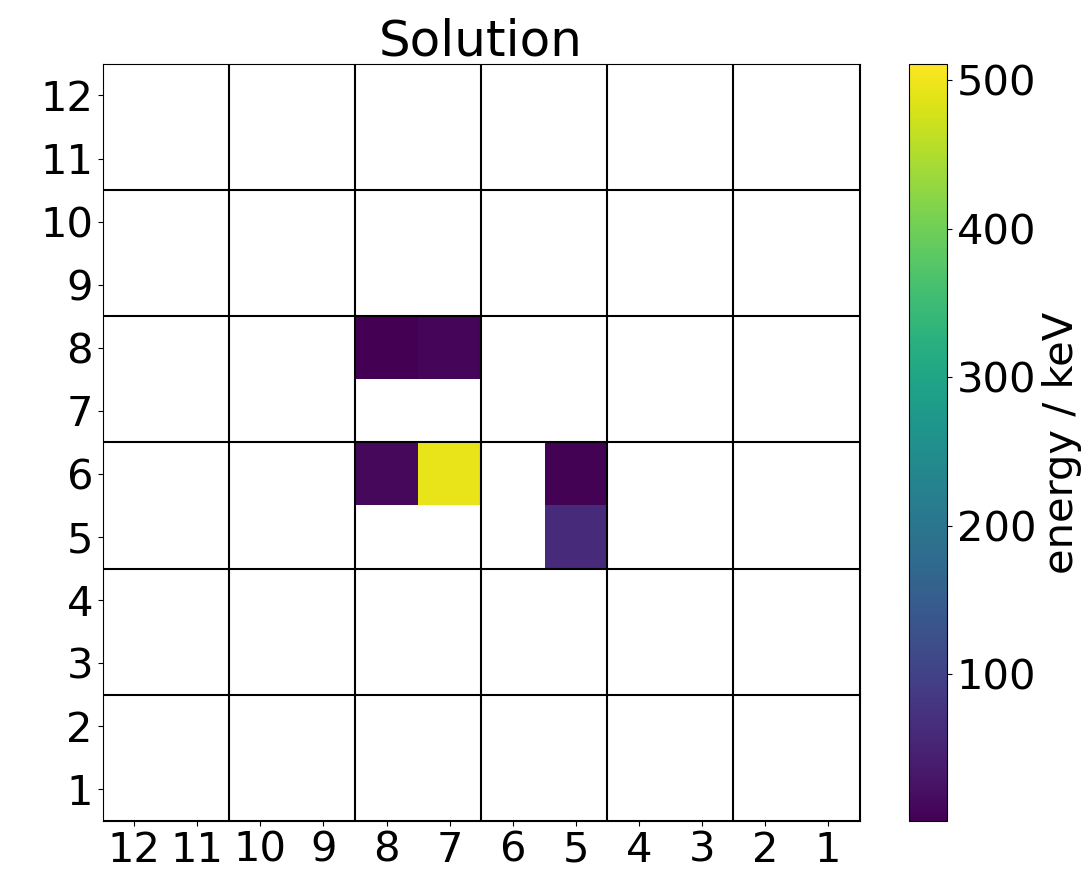}%
		\label{fig:algorithm_solution}}
	\caption{Crystal and energy distribution of one exemplary event for the simulation in (a), the simulation with added lightspread in (b) and the algorithm solution in (c). This example is one of the events with the worst metric values ($\Delta_{\textnormal{crystal}} = 0.21$ and $\Delta_{\textnormal{sum}} = 0.14$), which is caused by the high energy deposit of over $\SI{20}{\kilo\electronvolt}$ in crystal $(5,5)$. No energy filter is applied for the plots and the metrics given.}
	\label{fig:crystal_dist_map_event}
\end{figure}

\begin{figure}[h]
	\centering
	\subfloat[]{\includegraphics[width=0.3\textwidth]{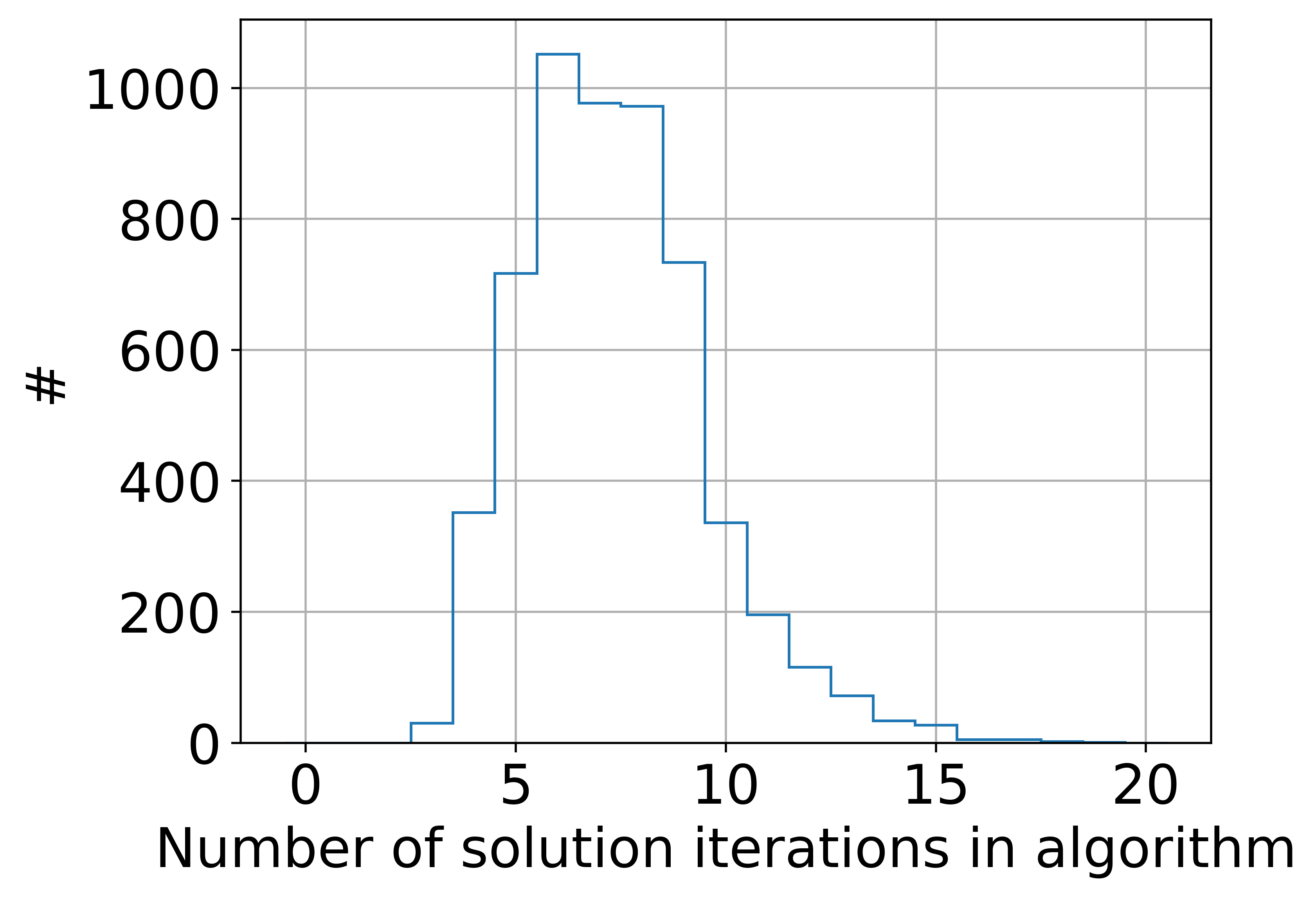}
		\label{fig:no_weighting}}
	\subfloat[]{\includegraphics[width=0.3\textwidth]{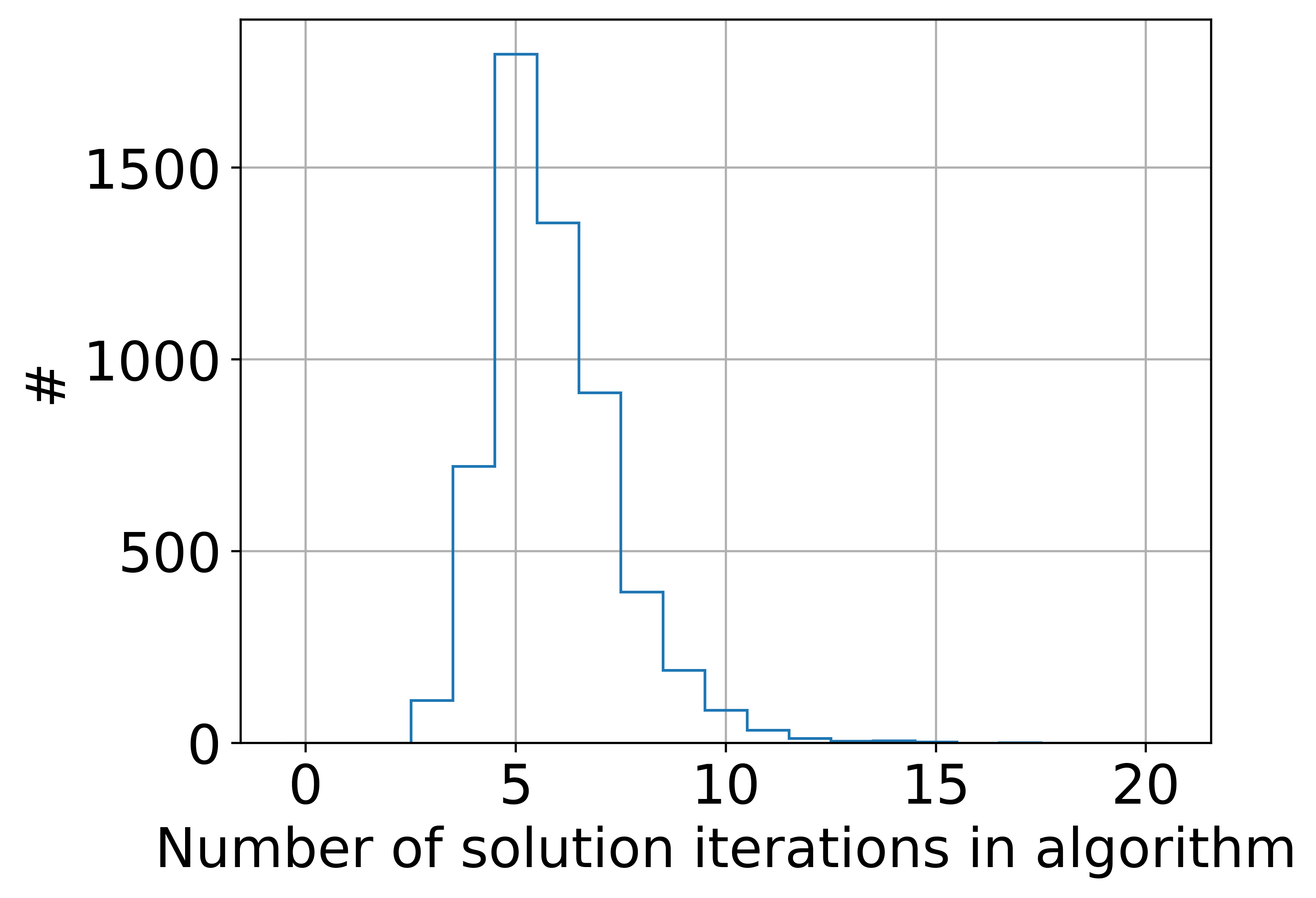}%
		\label{fig:photon_weighting}}
	\vspace{1mm}
	\subfloat[]{\includegraphics[width=0.3\textwidth]{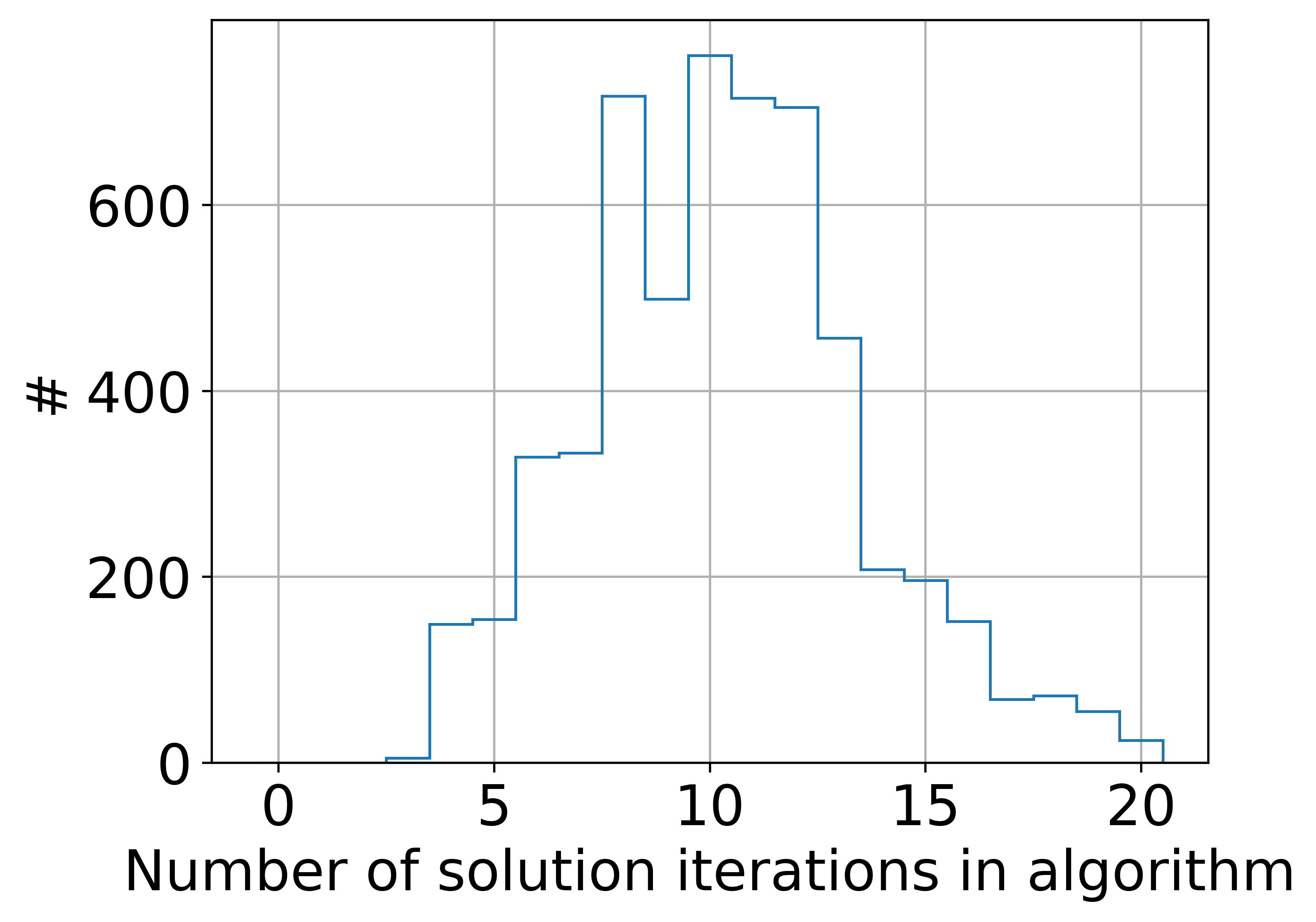}%
		\label{fig:rms_weighting}}
	\caption{Number of solving iterations within the algorithm for different weights. (a) shows no weights. In (b) photon weights were applied and in (c) the case for rms weights can be seen.}
	\label{fig:weighting}
\end{figure}

The first results of the validation show that the iterative reduction of the most negative energy contribution in the solving method results in many low energy contributions remaining in crystals not present in the simulation ground truth (compare Fig. \ref{fig:simulation} and \ref{fig:algorithm_solution}). These crystals also cause an overestimation of the event energy sum (see Fig. \ref{fig:energy_spectra}).\\ 
For the weighting approach, using the photon weights on both matrix and measurement broadened both metric distributions significantly and deteriorated the mean metric values (see Tab. \ref{tab:weighting_results}). The number of iterations the algorithm goes through is slightly reduced compared to the case without weighting (see Fig. \ref{fig:weighting}).  When applying the sigma weights to matrix and measurement, the $\Delta_{\textnormal{crystal}}$ distribution is shifted closer to zero and slightly sharpened, which is also shown in a smaller mean $\Delta_{\textnormal{crystal}}$ (see Tab. \ref{tab:weighting_results}). The behaviour for $\Delta_{\textnormal{sum}}$ is similar, reducing the mean metric value. The number of solving iterations in the algorithm seems to increase on average (see Fig. \ref{fig:weighting}). Also, the correct crystal fraction rises from $\SI{0.5}{\percent}$ (no weights) to $\SI{7.5}{\percent}$ (sigma weights) (see Tab. \ref{tab:weighting_results}).

\begin{figure}[h]
	\centering
	\includegraphics[width=0.6\textwidth]{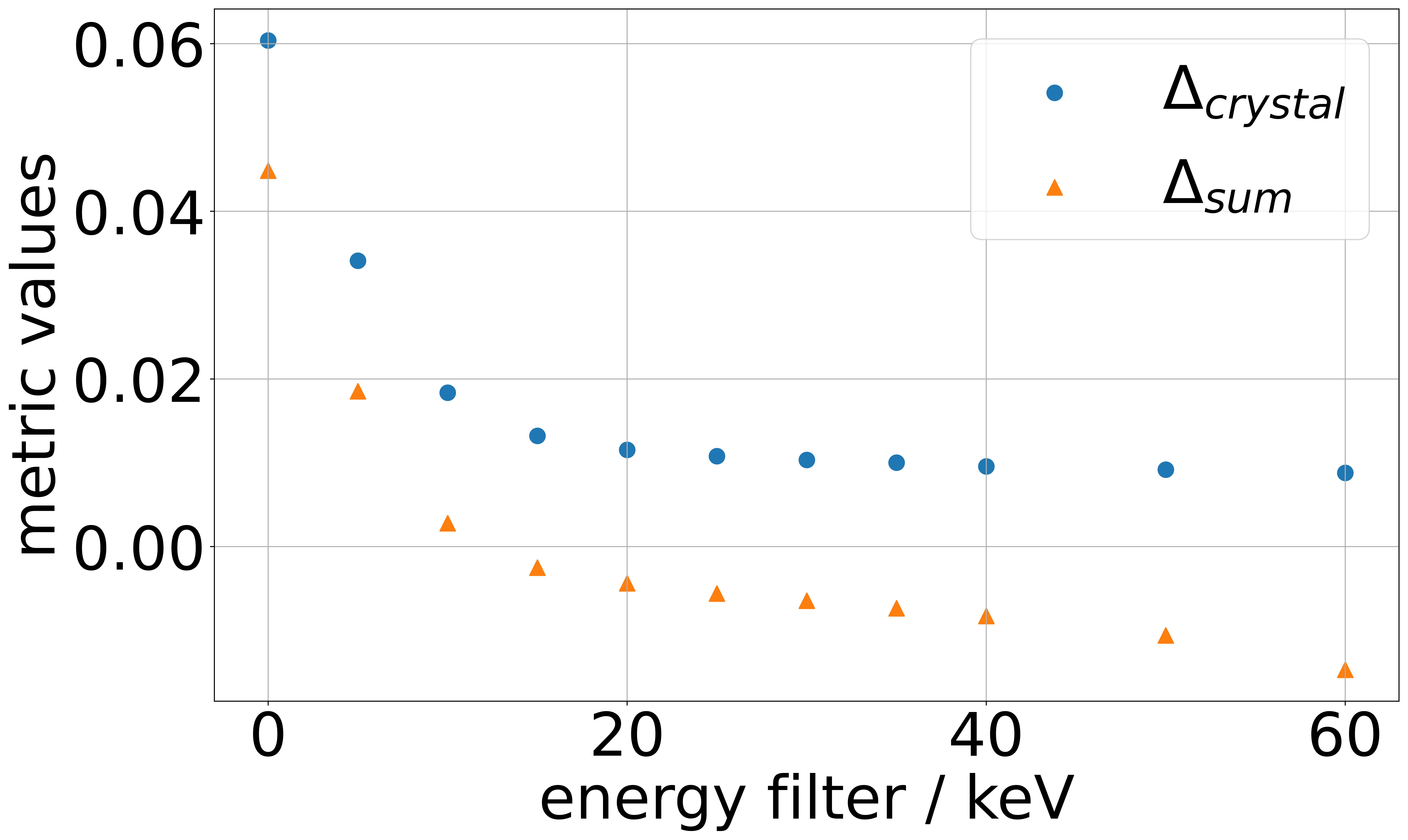}
	\caption{Metric values for different energy filter values are shown. The decrease of both metrics reduces around \SI{20}{\kilo\electronvolt}.}
	\label{fig:metric_vs_energyfilter}
\end{figure}

Using the posteriori filter without weighting and scanning through different values, we found that for a filter of $\SI{20}{\kilo\electronvolt}$, the $\Delta_{\textnormal{crystal}}$ metric goes into saturation (see Fig. \ref{fig:metric_vs_energyfilter}) and does not improve significantly for higher filter values, while the $\Delta_{\textnormal{sum}}$ metric crosses zero between filter values $\SI{10}{\kilo\electronvolt}$ and $\SI{15}{\kilo\electronvolt}$ and continues to deteriorate again from $\SI{20}{\kilo\electronvolt}$ on. Therefore, we chose an energy filter of $\SI{20}{\kilo\electronvolt}$ and applied it to all further evaluation with iterative reduction of solutions.

The energy filter removes most of the misidentified crystal solutions and reduces the energy overestimation significantly (see Fig. \ref{fig:energy_spectra}). With this filter, the correct crystal fraction rises from $\SI{0.5}{\percent}$ to $\SI{96.4}{\percent}$. It is also applied to both metrics (see Fig. \ref{fig:metric1_filtered20keV} and Fig. \ref{fig:metric2_filtered20keV}). Compared to the unfiltered case (see Fig. \ref{fig:metric1_unfiltered} and Fig. \ref{fig:metric2_unfiltered}), the metrics show a sharper distribution close to $0$ with only few outliers surpassing the absolute value of $0.05$. In general, $\SI{95.9}{\percent}$ of events fulfill this criterion. 
We found that combining sigma weighting and energy filter did not significantly improve the metric values compared to no weighting with energy filter and instead reduced the number of events with correctly identified crystals while increasing the average number of iterations within the algorithm. Therefore, we did not consider weights for further analysis, only using the energy filter instead.\\  

\begin{table*}[h]
	\centering
	{\caption{Mean metric values for different weighting and energy filtering combinations with iterative reduction of negative solutions. Additionally, the correct crystal fraction is shown.}
		\label{tab:weighting_results}}
	\setlength\tabcolsep{10pt}
		\begin{tabular}{@{}lS[table-format = 1.4(4), round-mode = places, round-precision = 4, table-figures-uncertainty=5, separate-uncertainty=true, table-align-uncertainty]S[table-format = 1.4(4), round-mode = places, round-precision = 4, table-figures-uncertainty=5]S[table-format = 3.0, round-integer-to-decimal, round-mode = places, round-precision = 1, table-figures-uncertainty=1]@{}}
			\toprule[1.5pt]
			\toprule
			& \multicolumn{1}{c}{$\overline{\Delta_{\textnormal{crystal}}}$} & \multicolumn{1}{c}{$\overline{\Delta_{\textnormal{sum}}}$} & \multicolumn{1}{c}{corr. crystal frac. / $\%$}\\ 
			\midrule[1.5pt]
			\textbf{no uncertainties} & & &\\
			\midrule[1pt]
			no weights & 0.0604(4) & 0.0449 \pm 0.0003 & 0.5\\
			\midrule
			photon weights & 0.2875 \pm 0.0027 & -0.1171 \pm 0.0021 & 0.0\\
			\midrule
			sigma weights & 0.0459 \pm 0.0006 & 0.0329 \pm 0.0006 & 7.5\\
			\midrule
			\makecell[l]{no weights\\ + posteriori filter \SI{20}{\kilo\electronvolt}} & 0.0120 \pm 0.0003 & -0.0044 \pm 0.0002 & 96.4\\
			\midrule
			\makecell[l]{sigma weights\\ + posteriori filter \SI{20}{\kilo\electronvolt}} & 0.0161 \pm 0.0005 & 0.0028 \pm 0.0005 & 89.4\\
			\midrule[1pt]
			\textbf{with uncertainties} & & &\\
			\midrule[1pt]
			\makecell[l]{no weights\\ + posteriori filter \SI{20}{\kilo\electronvolt}} & 0.0424 \pm 0.0005 & -0.0277 \pm 0.0007 & 96.6\\
			\midrule
			\makecell[l]{sigma weights\\ + posteriori filter \SI{20}{\kilo\electronvolt}} & 0.0421 \pm 0.0005 & -0.0212 \pm 0.0008 & 89.7\\
			\bottomrule
			\bottomrule[1.5pt]
	\end{tabular}
\end{table*}

The results with the energy filter led us to also test iteration methods with thresholds instead of iteratively reducing negative solutions. Removing all solutions below a set threshold, shows a quick improvement of metric values and fraction of events with correctly assigned crystals (see Tab. \ref{tab:threshold_results}) for increased thresholds.

\begin{table*}[h]
	\centering
	{\caption{Mean metric values for different thresholds during the solving iteration and no weighting. Additionally, the correct crystal fraction is shown.}
		\label{tab:threshold_results}}
	\setlength\tabcolsep{10pt}
	\begin{tabular}{@{}lS[table-format = 1.4(4), round-mode = places, round-precision = 4, table-figures-uncertainty=5, separate-uncertainty=true, table-align-uncertainty]S[table-format = 1.4(4), round-mode = places, round-precision = 4, table-figures-uncertainty=5, separate-uncertainty=true, table-align-uncertainty]S[table-format = 3.0, round-integer-to-decimal, round-mode = places, round-precision = 1, table-figures-uncertainty=1]@{}}
		\toprule[1.5pt]
		\toprule
		& \multicolumn{1}{c}{$\overline{\Delta_{\textnormal{crystal}}}$} & \multicolumn{1}{c}{$\overline{\Delta_{\textnormal{sum}}}$} & \multicolumn{1}{c}{corr. crystal frac. / $\%$}\\ 
		\midrule[1.5pt]
		\textbf{no uncertainties} & & &\\
		\midrule[1pt]
		threshold \SI{0}{\kilo\electronvolt} & 0.0441 \pm 0.0004 & 0.0309 \pm 0.0003 & 3.6\\
		\midrule
		threshold \SI{5}{\kilo\electronvolt} & 0.0101 \pm 0.0002 & 0.0008 \pm 0.0002 & 90.2\\
		\midrule
		threshold \SI{10}{\kilo\electronvolt} & 0.0092 \pm 0.0002 & 0.0001 \pm 0.0002 & 95.0\\
		\midrule
		threshold \SI{20}{\kilo\electronvolt} & 0.0101 \pm 0.0003 & -0.0004 \pm 0.0002 & 95.1\\
		\midrule[1pt]
		\textbf{with uncertainties} & & &\\
		\midrule[1pt]
		threshold \SI{10}{\kilo\electronvolt} & 0.0529 \pm 0.0010& -0.0356 \pm 0.0011 & 95.4\\
		\midrule
		threshold \SI{20}{\kilo\electronvolt} & 0.0541 \pm 0.0011 & -0.0359 \pm 0.0011& 93.6\\
		\bottomrule
		\bottomrule[1.5pt]
	\end{tabular}
\end{table*}

\subsection{Simulation with uncertainties}
\begin{figure}[h]
	\centering
	\begin{minipage}[c]{0.49\textwidth}
		\centering
		\subfloat[]{\includegraphics[scale=0.5]{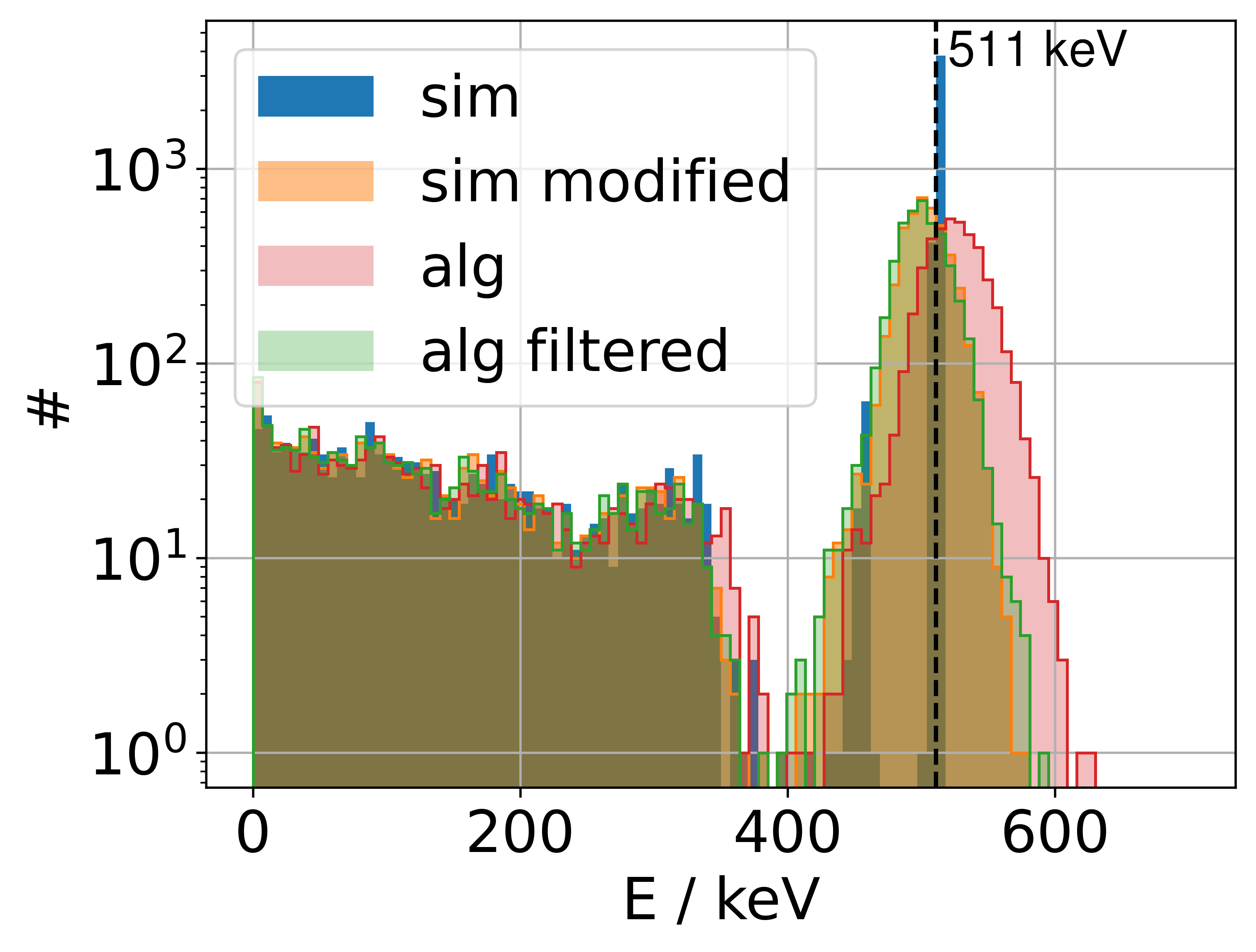}%
			\label{fig:broadened_energy_spectra}}
	\end{minipage}%
	\hspace{0mm}
	\begin{minipage}[c]{0.49\textwidth}
		\centering
		\subfloat[]{\includegraphics[scale=0.4]{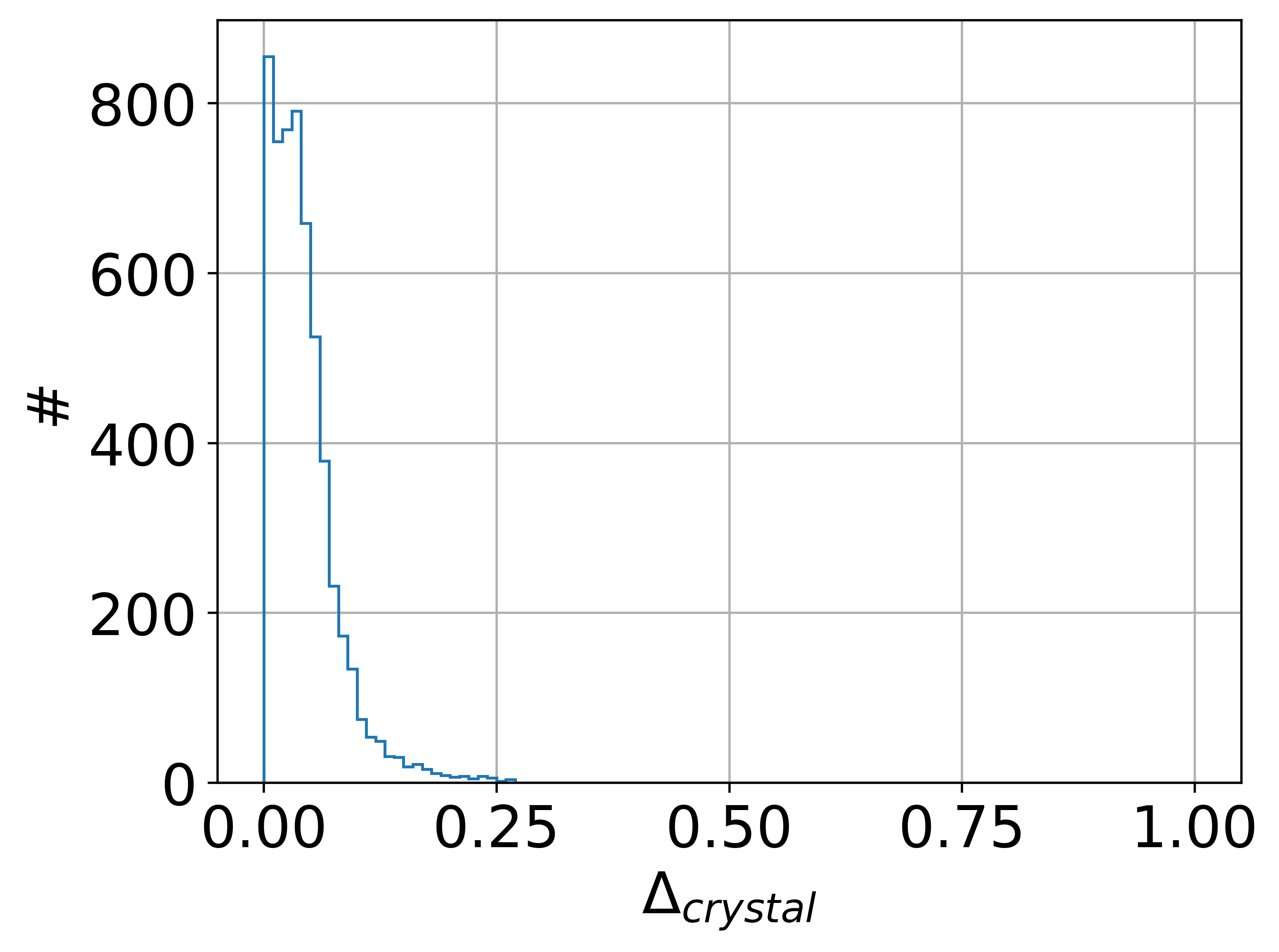}
		\label{fig:broadened_metric1_filtered20keV}}
		\vfill
		\subfloat[]{\includegraphics[scale=0.4]{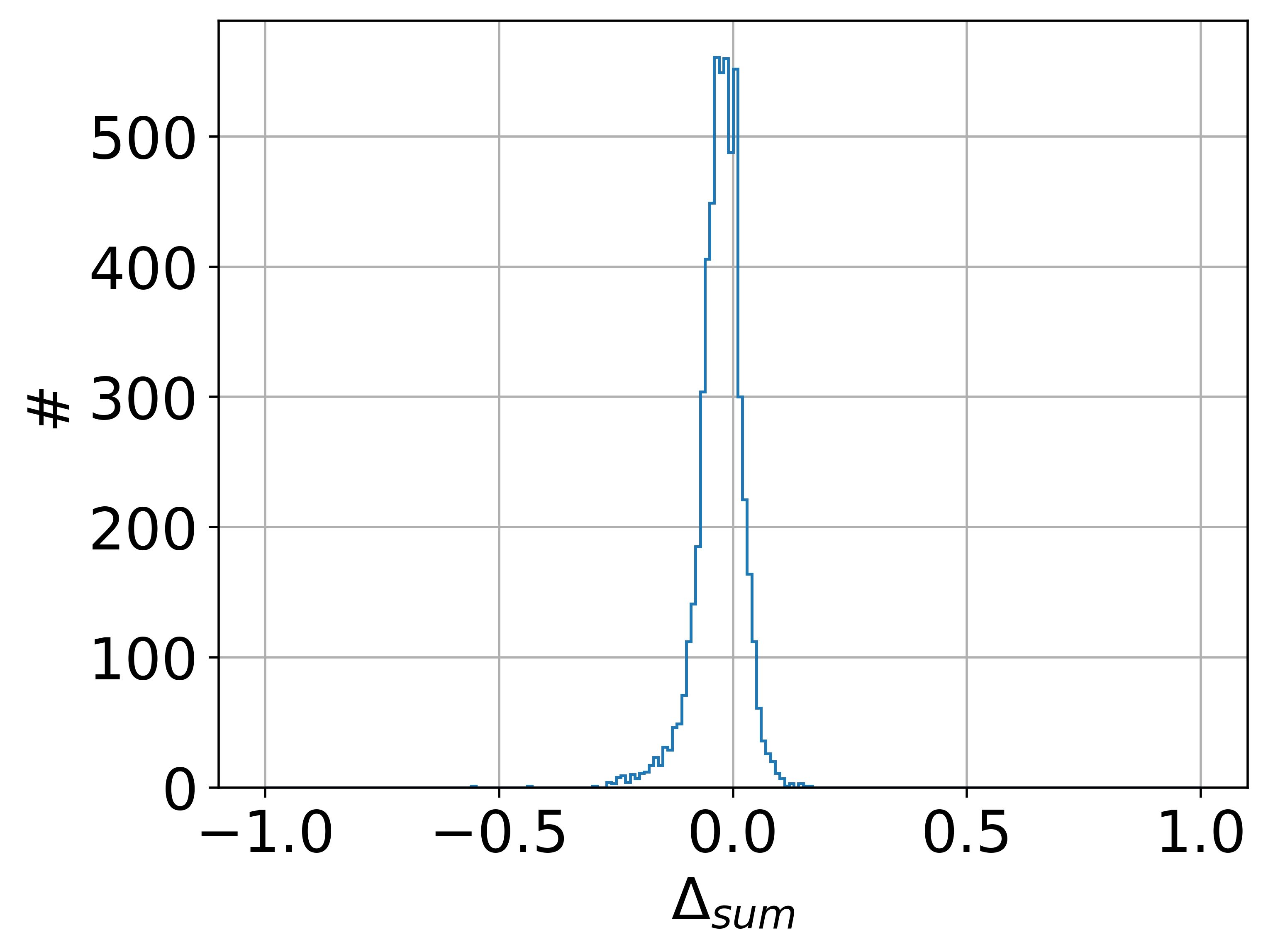}
		\label{fig:broadened_metric2_filtered20keV}}
	\end{minipage}%

	\vspace{0mm}

	\caption{In (a) the event energy sum spectra of the simulation (sim), the simulation with uncertainties (sim modified), the algorithm solution (alg) and the algorithm solution with an energy filter of \SI{20}{\kilo\electronvolt} (alg filtered) are shown. The algorithm solution spectrum shows a structure similar to the simulation with uncertainties, but slightly overestimates the energy. If the energy filter is applied, this effect is significantly reduced (alg filtered) and modified simulation and filtered algorithm solution fit together satisfactorily. (b) demonstrates the distribution of the $\Delta_{\textnormal{crystal}}$ metric with applied \SI{20}{\kilo\electronvolt} energy filter, while (c) shows the same for the $\Delta_{\textnormal{sum}}$ metric.}
	\label{fig:results_broadened}
\end{figure}

After including the uncertainties described in section \ref{sec:materials}, the energy spectrum of the simulation is broadened and shifted slightly towards lower energies (see Fig. \ref{fig:broadened_energy_spectra}). The algorithm solution with a posteriori filter of \SI{20}{\kilo\electronvolt} reproduces the shape and position of the modified simulation's energy spectrum. Both metrics cluster mainly around zero with a broader distribution than before modifying the simulation (see Fig. \ref{fig:broadened_metric1_filtered20keV} and Fig. \ref{fig:broadened_metric2_filtered20keV}). The $\Delta_{\textnormal{sum}}$ metric is also slightly shifted towards negative values. Applying the sigma weights to this modified dataset showed no significant improvement of the mean and a slight decrease in the correct crystal fraction.
\label{sec:results}

\section{Discussion}
\begin{figure}[h]
	\centering
	\subfloat[]{\includegraphics[width=0.25\textwidth]{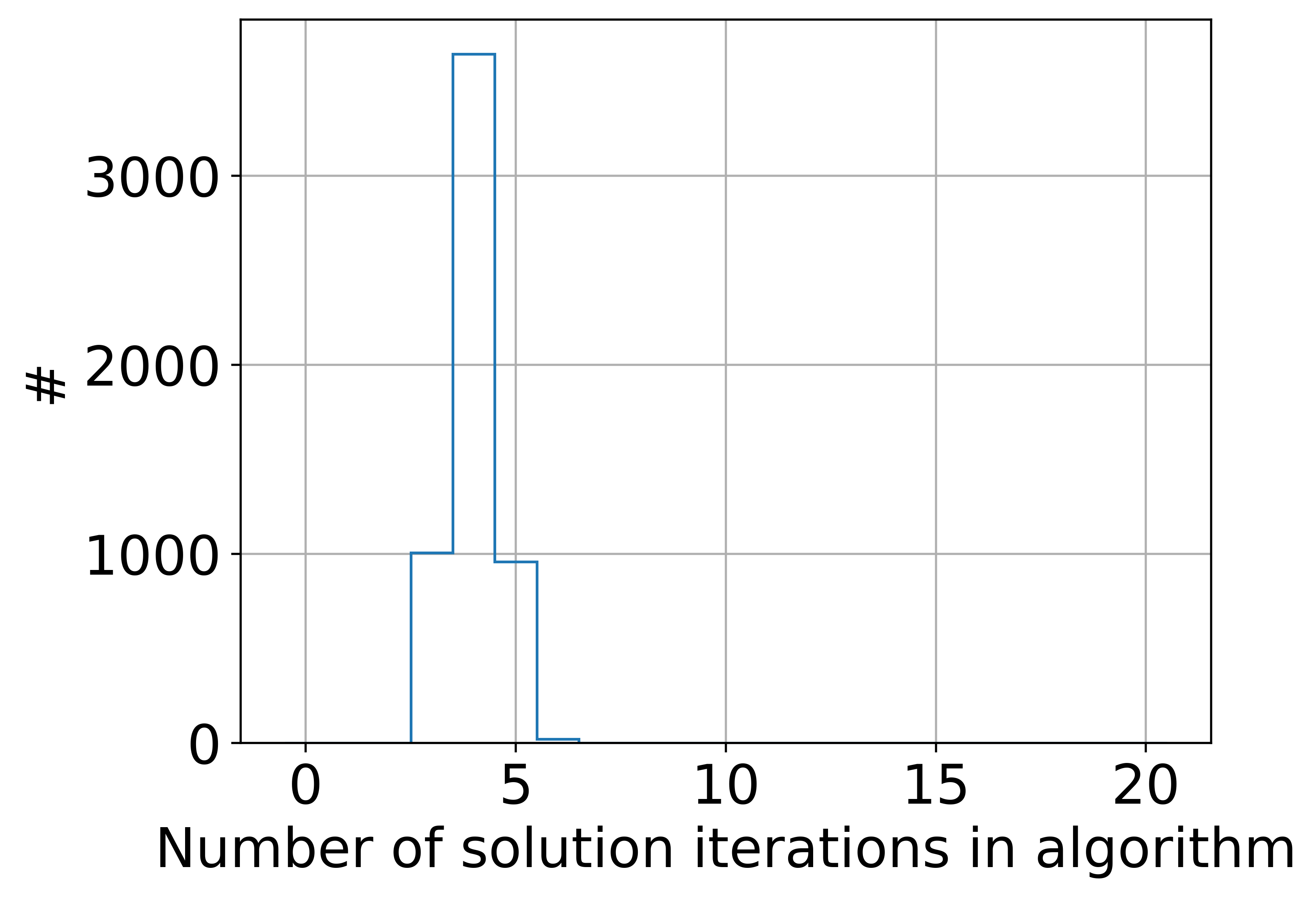}
		\label{fig:thres0}}
	\subfloat[]{\includegraphics[width=0.25\textwidth]{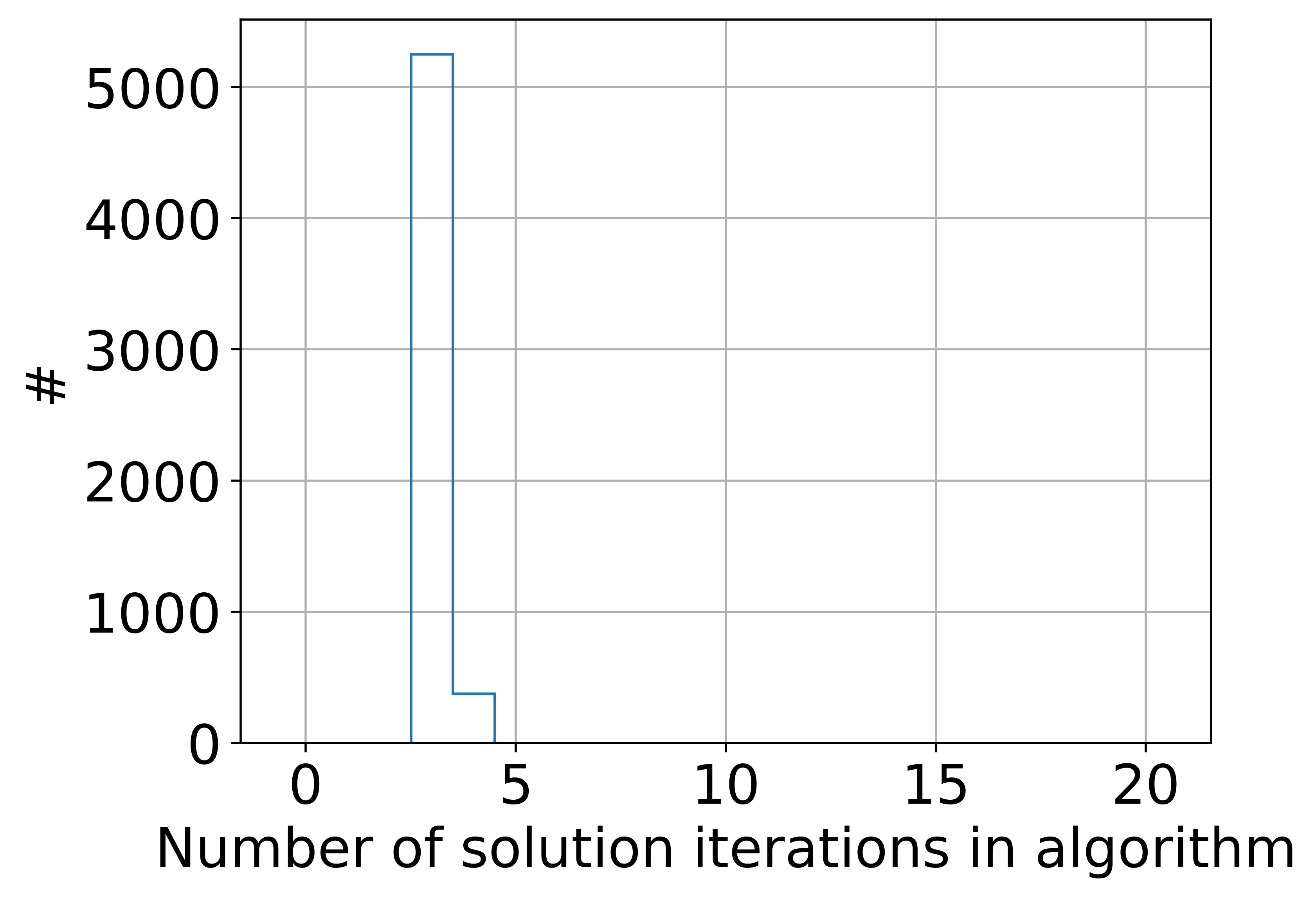}%
		\label{fig:thres5}}
	\subfloat[]{\includegraphics[width=0.25\textwidth]{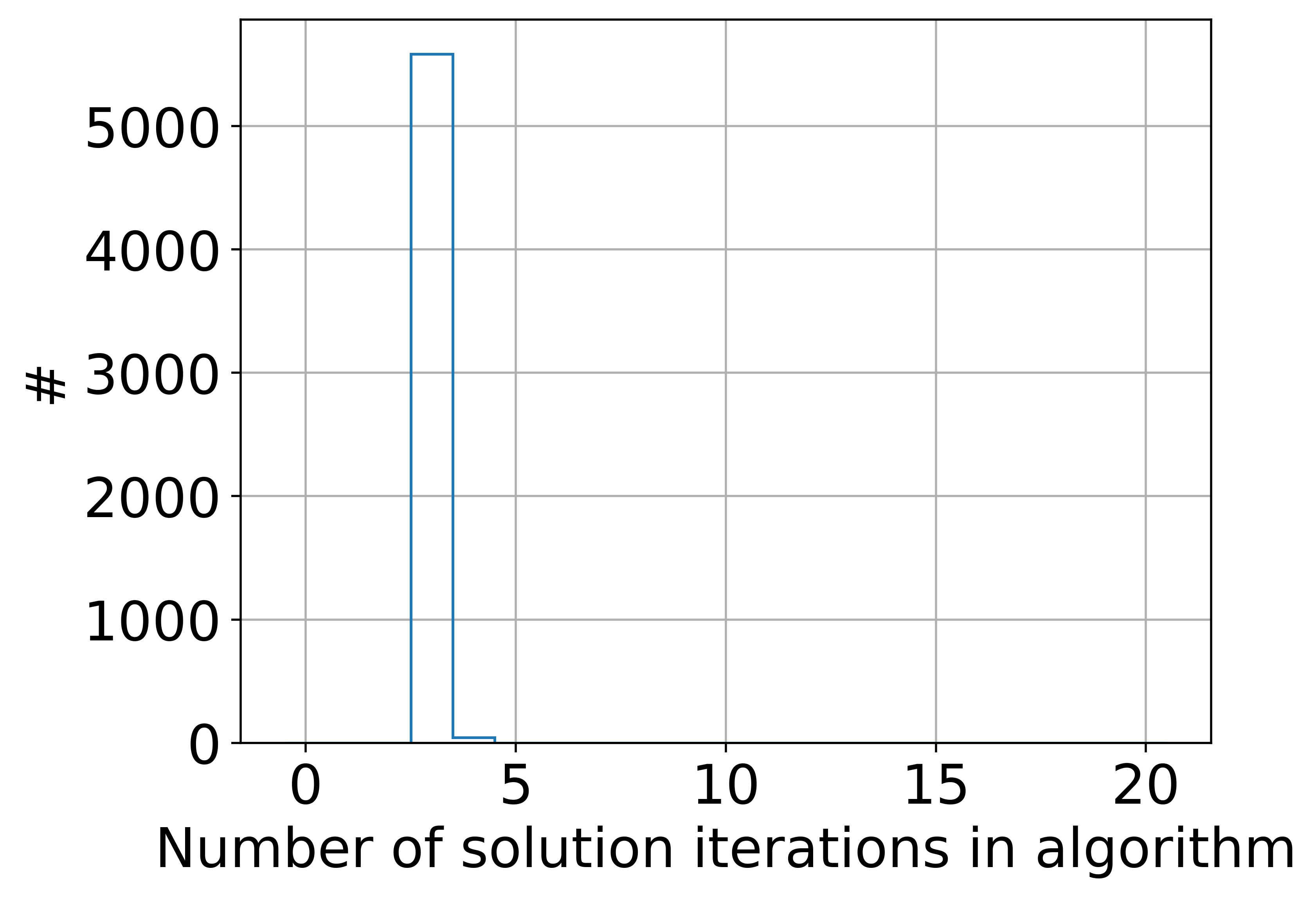}%
		\label{fig:thres10}}
	\subfloat[]{\includegraphics[width=0.25\textwidth]{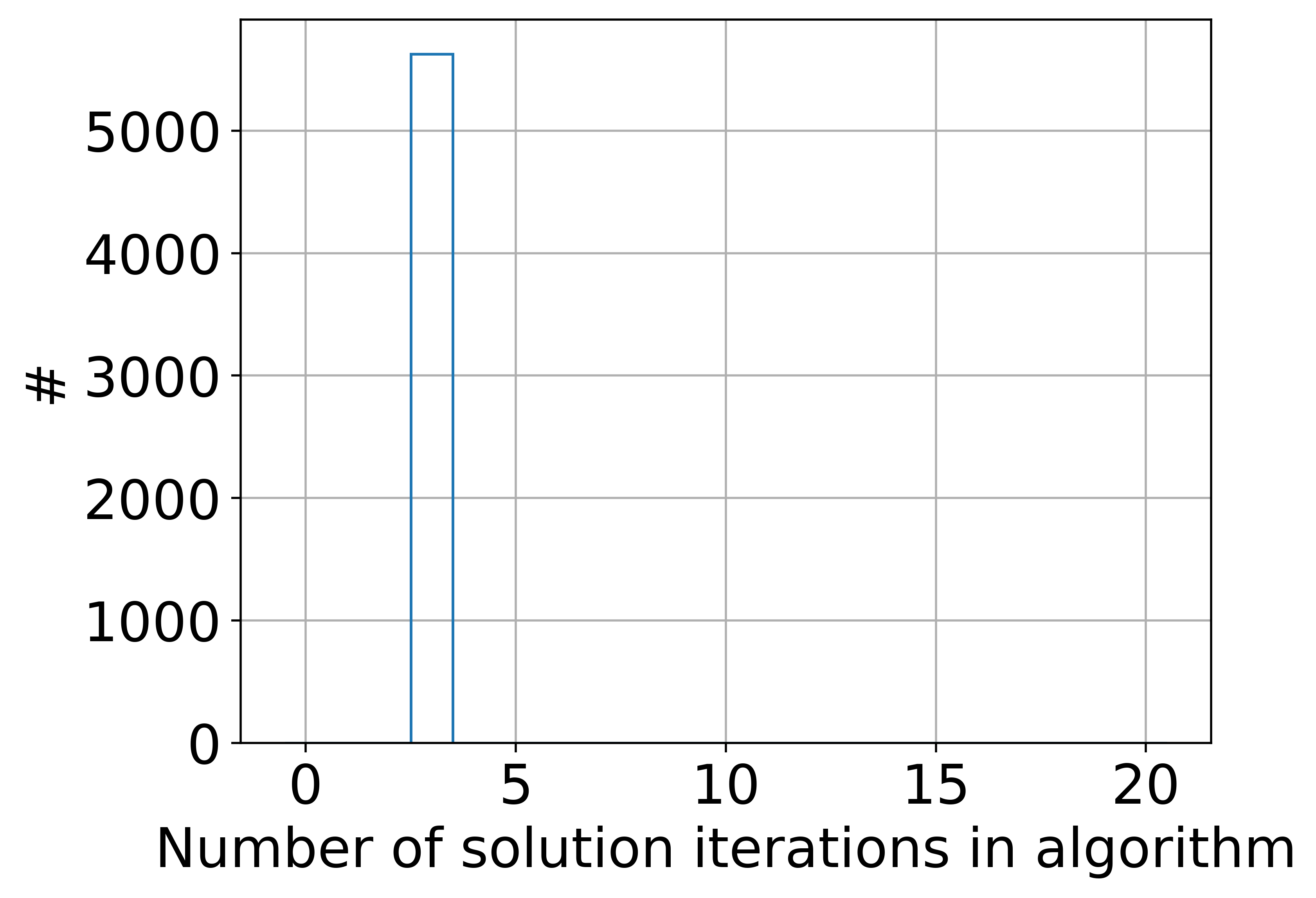}%
		\label{fig:thres20}}
	\caption{Number of solving iterations within the algorithm for different weights. (a) shows a threshold of \SI{0}{\kilo\electronvolt}. In (b) a threshold of \SI{5}{\kilo\electronvolt} is applied, (c) displays a threshold of \SI{10}{\kilo\electronvolt} and in (d) the case of a \SI{20}{\kilo\electronvolt} threshold is shown.}
	\label{fig:thresholds}
\end{figure}

The overall validation of the described algorithm is deemed successful, with a high correct crystal fraction after applying the posteriori filter, and a low deviation of energy contributions in these crystals as described by low metric values. The correct crystal fraction of more than \SI{96}{\percent} can be roughly compared to the identification rate stated by Lee et al (2018), which lies at \SI{93}{\percent} for one-to-one coupled detectors \cite{Lee2018}, showing that our algorithm can indeed hold up to previously published identification rates. With our defined metrics, we have given an estimate on the accuracy of energies reproduced by the algorithm. According to $\Delta_\textnormal{sum}$ the event energies were on average reproduced with a deviation of \SI{0.28}{\percent} and more than \SI{99}{\percent} of all simulated events showed an energy deviation of less than \SI{0.5}{\percent}. With this result, like Lee et al \cite{Lee2018}, we can claim a good linearity between recovered and true energy. Comparing to ICS recovery with a convolutional neural network \cite{Lee2021}, where a crystal selection accuracy of \SI{99}{\percent} was stated for photoelectric effect and \SI{91}{\percent} for ICS events, our correct crystal fraction of \SI{96}{\percent} encompasses both of these values and tests in more detail whether a correct crystal recovery was made, which is why we believe that it should be considered on the same level or even improved. As the energy linearity in \cite{Lee2021} is similar to that in \cite{Lee2018}, we can assume the same conclusion as before, that our energy recovery is on a similar level as the convolutional neural network.\\
The tests to apply different weighting during the recovery process show that for simulations without uncertainties the mean metric values are deteriorated for photon weights and slightly improved for sigma weights. A disadvantage of the sigma weights is that they increase the number of iterations, which the algorithm goes through. The improvement with the sigma weights is, however, outweighed by the application of an energy filter. Even a filter of \SI{5}{\kilo\electronvolt} already achieves better results than the sigma weights (see Fig. \ref{fig:metric_vs_energyfilter}). Using the optimal energy filter value of \SI{20}{\kilo\electronvolt} and no weights, achieves the best mean metric values and the best fraction of correctly assigned events (see Tab. \ref{tab:weighting_results}). We have also found that the combination of sigma weights and energy filter has no significant effect on the metric values, while deteriorating the fraction of correctly assigned events.\\
Applying uncertainties to the simulation, the algorithm can more easily reproduce the shape of the energy spectrum and with the energy filter manages to replicate also the position of the peak very well (see Fig. \ref{fig:broadened_energy_spectra}). The mean metric values of the simulation data with uncertainties and energy filter are similar to those with no uncertainties and no energy filter.\\
If, for simulations without uncertainties, we compare the mean metric values of the algorithm with a threshold in the iterative solving process and the algorithm with iteratively removed negative solutions and posteriori energy filter of \SI{20}{\kilo\electronvolt}, it becomes evident that even a small threshold of \SI{5}{\kilo\electronvolt} makes both metric values comparable.
Only the correct crystal fraction is slightly reduced, but this effect can be removed by increasing the threshold to \SI{10}{\kilo\electronvolt} or higher. Applying these higher thresholds on the simulation with uncertainties shows again similar results to those for application of the posteriori filter, but overall they are slightly deteriorated. At the same time, the use of thresholds reduces the mean and maximum number of solving iterations significantly (see Fig. \ref{fig:weighting} and Fig. \ref{fig:thresholds}).\\
Filtering low energy crystal contribution, either by thresholding or by applying an external filter, can be considered a viable procedure in this case, as the electronics in the experiment impose a low energy cut-off on channel basis through rising edge thresholds or digital validation schemes, such as described in Tabacchini et al (2014) \cite{Tabacchini2014} and energy thresholds are commonly used in reconstruction to filter Compton scatter and with that also ICS \cite{Rechka2009}. Of course, the experimental argument is only valid if we are considering a one-to-one coupled detector block. For the more complicated case of light-sharing detectors, the method would have to be reevaluated. 
Considering the application of this algorithm with a phantom or patient as the gamma source, we believe that the algorithm would be capable of also recovering phantom or patient scatter events and ICS events related to them. The algorithm described in this work does not make any a-priori estimation regarding the energy of the incoming gamma. It primarily considers the measurement. The data on which the algorithm is applied is reduced to the channels, which have photon values larger zero and all crystals, which could contribute to a light output on these channels (see section \ref{sec:algorithm}). In this regard, a gamma, scattered in the phantom or patient, should not be different from a gamma scattered in the scintillator, where the scattered gamma is not stopped in the crystal. The latter case is covered in our simulation validation. Therefore, the algorithm should not distinguish between gamma photons scattered in the phantom or patient and those which do not interact with either. Thus, it should be able to recover the crystals of both events sufficiently.

\label{sec:discussion}

\section{Conclusion and Outlook}
In summary, we have developed and validated an algorithm, which recovers individual crystals from a gamma photon interaction with their deposited energies. The algorithm uses a measurement-based calibration in combination with a numerical least-square solver for the recovery. From the validation results we can conclude, that the algorithm can reproduce the events of the simulation satisfactorily. For a simulation without uncertainties, we have shown that up to $\SI{96.4}{\percent}$ of solution crystals with energies above $\SI{20}{\kilo\electronvolt}$ correspond to the simulation crystals and $\SI{95.9}{\percent}$ of events have a total energy deviation of less than $\SI{5}{\percent}$ between simulation and algorithm. The application of weights to the LSM and measurement did not yield a significant performance improvement in combination with the mentioned energy filter. Similarly, using different methods for the iteration process did not show a difference in performance, but thresholding displayed a reduction in iteration numbers, making it favourable, especially for high data throughput. The performance of the algorithm holds for simulations with uncertainties.\\
We see great potential to use the crystal interaction information to improve efficiency and signal-to-noise ratio of the PET reconstruction. Based on the crystal interaction information, Compton kinematics together with an incidence angle would allow to determine the most likely first interaction crystal. Passing on the most probable first interaction crystal would be a benefit to the reconstruction, as demonstrated by Surti et al (2018) \cite{Surti2018}, who showed that choosing the crystal with the second highest energy deposit improved positioning performance compared to using the crystal with the highest energy deposit. With our algorithm the second highest crystal method would not be limited to one-to-one coupled detectors anymore. Therefore, another future aim is to apply the developed algorithm to highly pixelated detectors and prove its efficiency. Additionally, it is planned to implement the algorithm for the PET-MRI system recently installed in University Medical Center Utrecht \cite{Utrecht2021}.
\label{sec:conclusion}

\section*{Acknowledgments}
This work was carried out with funding from the JARA Prep Fund Simulation and Data Science "Deep Image Data Analysis for Precision Medical Imaging" (PF-JARA-SDS008). 

\section*{Conflict of Interest}
The authors declare the following financial interests/personal relationships which may be considered as potential competing interests with the work reported in this paper: V.S. and D.S. are co-founders and employees of the spin-off company Hyperion Hybrid Imaging Systems GmbH, Aachen, Germany.
\bibliographystyle{./medphy.bst}
\bibliography{abstract}

\end{document}